\documentclass[]{rsproca_new}

\usepackage{overpic}
\usepackage{psfrag}
\usepackage[normalem]{ulem}

\begin{document}

\title{
Robust identification of harmonic oscillator parameters using the adjoint Fokker-Planck equation
}

\author{
E. Boujo and N. Noiray
}

\address{CAPS Laboratory, 
   Mechanical and Process Engineering Department, 
   ETH Z\"urich, 
   Switzerland
}

\subject{Mathematical physics}

\keywords{
System identification, 
Langevin equation, 
Fokker-Planck equation,
Adjoint methods
}

\corres{
E. Boujo\\
\email{eboujo@ethz.ch}\\
N. Noiray\\
\email{noirayn@ethz.ch}
}

\begin{abstract}

We present a model-based output-only method for identifying from time series the parameters governing the dynamics of stochastically forced oscillators. In this context, suitable models of the oscillator's damping and stiffness properties are postulated, guided by physical understanding of the oscillatory phenomena. The temporal dynamics and the probability density function of the oscillation amplitude are  described by a Langevin equation and its associated Fokker-Planck equation, respectively. One method consists in fitting the postulated analytical drift and diffusion coefficients with their estimated values, obtained from data processing by taking the short-time limit of the first two transition moments. However, this limit estimation loses robustness in some situations - for instance when the data is band-pass filtered to isolate the spectral contents of the oscillatory phenomena of interest.
In this paper, we use a robust alternative where the adjoint Fokker-Planck equation is solved to compute Kramers-Moyal coefficients exactly, and an iterative optimisation yields the parameters that best fit the observed statistics simultaneously in a wide range of amplitudes and time scales. The method is illustrated with a stochastic Van der Pol oscillator serving as a prototypical model of thermoacoustic instabilities in practical combustors, where system identification is highly relevant to control.

\end{abstract}

\begin{fmtext}

\end{fmtext}

\maketitle

\graphicspath{{./}}

\clearpage
\newpage
\section{Introduction}

\begin{figure}[!h] 
\psfrag{f}[t][]{$\omega$} 
\psfrag{g}[][]{} 
\psfrag{ppp}[l][l]{\scriptsize $\,\eta$} 
\psfrag{pp1}[l][l]{\scriptsize $\,\eta_1$} 
\psfrag{pp2}[l][l]{\scriptsize $\,\eta_2$} 
\psfrag{pp3}[l][l]{\scriptsize $\,\eta_3$} 
\psfrag{PP}[l][l]{\scriptsize }
\psfrag{p1}[l][l]{\scriptsize } 
\psfrag{p2}[l][l]{\scriptsize } 
\psfrag{p3}[l][l]{\scriptsize } 
\psfrag{Pp1}[][]{\scriptsize \textcolor[rgb]{0,0.5,0}{$P_\infty(\eta_1)$}}
\psfrag{Pp2}[][]{\scriptsize \textcolor[rgb]{0,0,1}{$P_\infty(\eta_2)$}}
\psfrag{Pp3}[][]{\scriptsize \textcolor[rgb]{1,0,0}{$P_\infty(\eta_3)$}} 
\psfrag{S}[b][]{$\mathcal{S}_{\eta\eta} / \max(\mathcal{S}_{\eta\eta})$ [dB]}
\centerline{  
   \hspace{0.8cm}
   \begin{overpic}[height=5.5cm,tics=10]{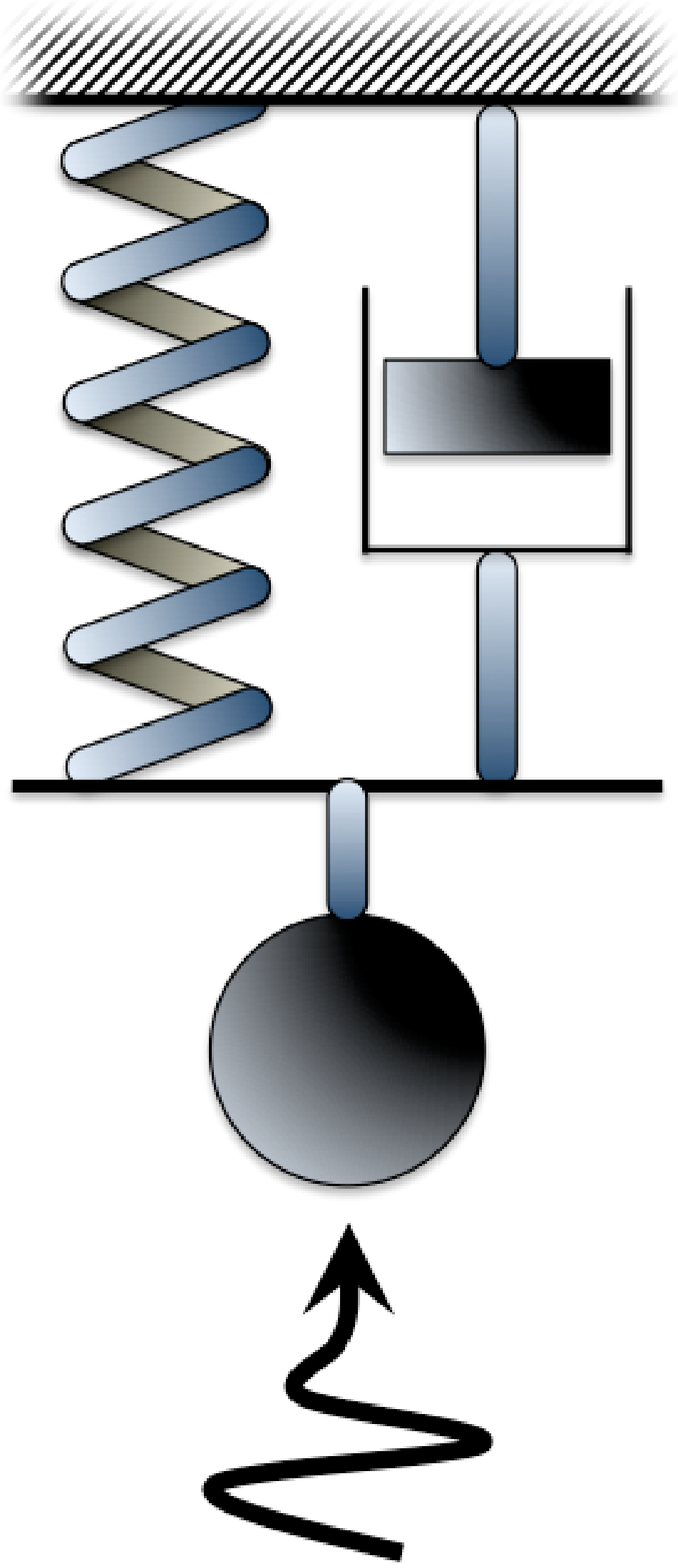}
      \put(-18,95){$(a)$}
      \put(-13,74){\scriptsize linear}
      \put(-15,68){\scriptsize stiffness}
      \put( 44,74){\scriptsize nonlinear}
      \put( 45,68){\scriptsize damping}
      \put( 35,13){\scriptsize external}
      \put( 36, 7){\scriptsize forcing}
   \end{overpic} 
   \hspace{2.5cm}
   \begin{overpic}[height=5.5cm,tics=10]{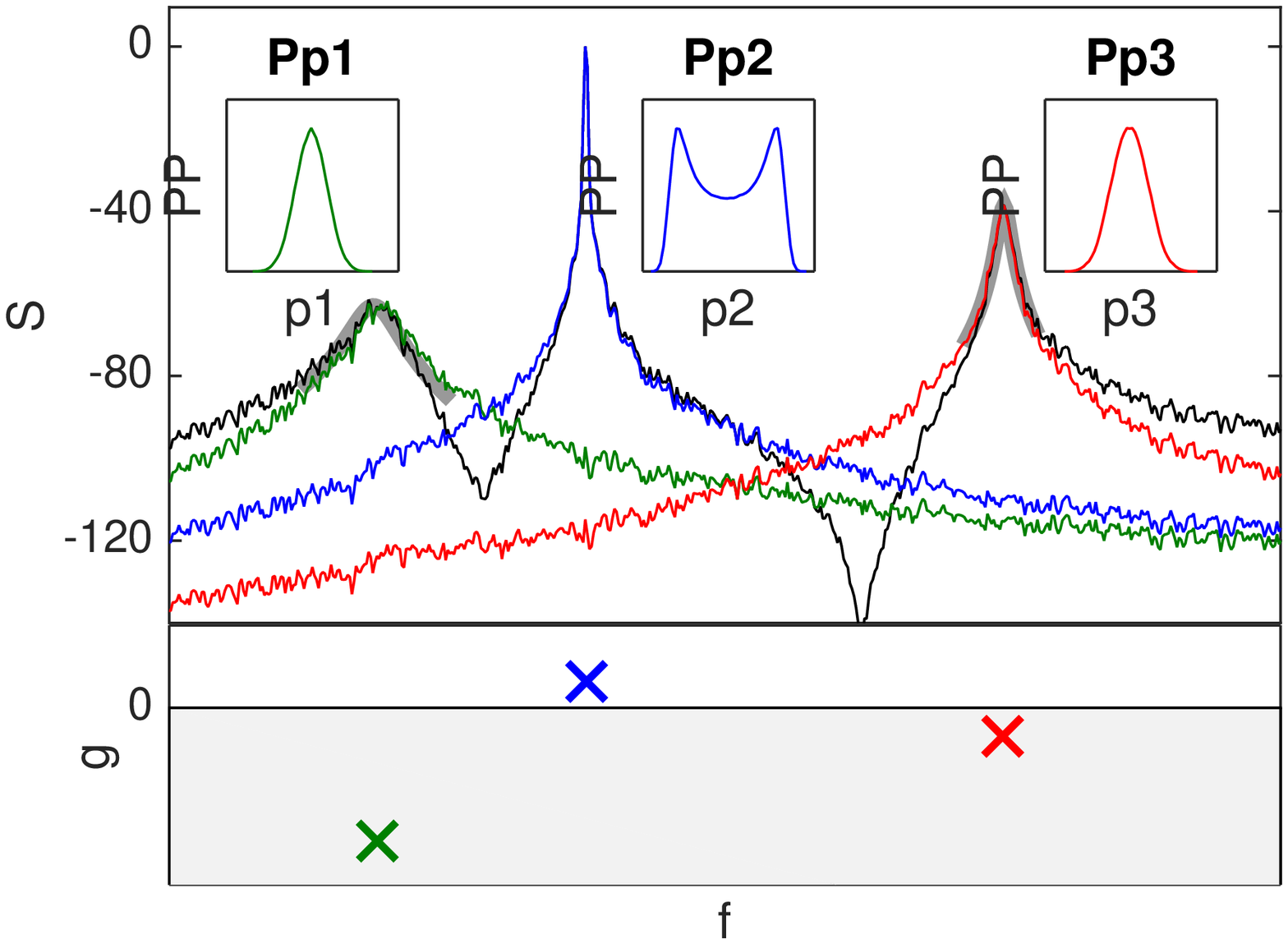}
      \put(-12,70){$(b)$}
      \put( -2, 8){\rotatebox{90}{growth}}
      \put(  2,11){\rotatebox{90}{rate}}
      \put( 85,21){\scriptsize unstable}
      \put( 89,15){\scriptsize stable}
      \put( 20,38){\scriptsize \textcolor[rgb]{0,0.5,0}{$\eta_1$}}
      \put( 25,33){\scriptsize  \textcolor[rgb]{0,0,1}{$\eta_2$}}
      \put( 30,28){\scriptsize  \textcolor[rgb]{1,0,0}{$\eta_3$}}
      \put( 51,28){\scriptsize $\eta = \sum \eta_j$}
   \end{overpic} 
}
\caption{
$(a)$~Archetypal example of mechanical harmonic oscillator driven by external forcing: mass-spring-damper system. In this paper, the focus will be on systems with linear stiffness and nonlinear damping.
$(b)$~Typical frequency spectrum (power spectral density, upper panel) and corresponding poles (lower panel) of a system made of three coupled harmonic oscillators.
Thick lines: Lorentzian fits for the two linearly stable modes.
Insets: stationary probability density functions.
} 
\label{fig:intro}
\end{figure}

Harmonic oscillators are ubiquitous in nature as well as in technological applications. In many cases, the oscillators are subject to random noise forcing. This is a wide topic, highly relevant to a broad range of phenomena in domains ranging across mechanics, physics, chemistry, electronics, biology, sociology and virtually all fields~\cite{Moshinsky96}. 
Harmonic oscillators can be described in general by the second-order differential equation
\begin{equation}
\ddot \eta  + \omega_{0}^2 \eta =  f(\eta,\dot\eta) + g(t),
\label{eq:eta0}
\end{equation}
where $\omega_{0}/2\pi$ is the natural frequency,
$f(\eta,\dot\eta)$ is a nonlinear function of the state and its derivative, and $g(t)$ is an external time-dependent forcing which might be deterministic or stochastic (see for example figure~\ref{fig:intro}$a$).
The linear stability properties of a deterministic, unforced oscillator depend on the linear terms of $f(\eta,\dot\eta)$: the equilibrium becomes linearly unstable if the linear damping (proportional to $\dot\eta$) is negative, while the linear stiffness (proportional to $\eta$) affects the oscillation frequency. Nonlinearities induce a variety of interesting phenomena, such as supercritical and subcritical bifurcations,  bistability, hysteresis, and chaos.
Equation (\ref{eq:eta0}) has been studied extensively and the system's behaviour is well understood
both in the linearly stable and unstable regimes, either without forcing or with forcing of various types: deterministic forcing (e.g. $g(t)$ is  a step, an impulse, or a time-harmonic function), 
stochastic additive forcing (external noise induces a random force  $g(t)$) or stochastic multiplicative forcing (external noise induces random fluctuations of the coefficients in $f(\eta,\dot\eta)$) \cite{Gitterman2005,Gitterman2012}. Coupling between several oscillators introduces another layer of complex and rich phenomena such as synchronisation~\cite{Pikovsky01,Balanov09}.

Although there has been much progress in understanding and  characterising the behaviour of oscillators with known parameters, the inverse problem of system identification (SI), i.e. finding the unknown parameters of a given system, is a challenging task. In many situations, one cannot study the system's response to a prescribed external input: applying such a forcing is not practical due to the scales involved or due to the large power needed (e.g. climate oscillations such as fluctuations in  atmospheric and oceanic temperatures~\cite{Baldwin01,Dijkstra02}; stellar pulsations~\cite{Cox80,Brown94}; molecular vibrations~\cite{LandauLifshitz69}; pressure oscillations in high energy density systems such as gas turbines, aero- and rocket engines).
However, one can take advantage of the system being driven by inherent stochastic forcing. For instance, in the linearly stable regime, measuring the power spectral density of an oscillator subject to additive noise allows the identification of the linear parameters, as illustrated in figure~\ref{fig:intro}$b$: the system is composed of three weakly coupled nonlinear oscillators; two of the corresponding modes are linearly stable, thus a Lorentzian fit of the frequency spectrum around the peak frequency yields a good measure of the (negative) linear growth rate and of the noise intensity.
Unfortunately, this simple method does not work in the unstable regime, neither does it give access to nonlinear parameters.

The difficulty can be circumvented by turning to output-only SI, i.e. measuring  one or several observables and analysing the data. One such well-known SI technique is time series analysis, that aims at reproducing and predicting time series, using for instance autoregressive models where the current state of the system depends on past states, and where random fluctuations are treated as a measurement noise (that does not affect the system's dynamics) \cite{Hamilton94,Shumway11}.
Alternatively, treating fluctuations as a dynamic noise (that affects the  dynamics) and adopting a statistical viewpoint proves particularly convenient: the deterministic properties of a system are related to (and can therefore be extracted from) the statistical properties of stochastic time series.
Specifically, this kind of analysis is based on the Fokker-Planck (FP) equation, which governs the evolution of  probability density functions. 
Under some assumptions, the FP equation is associated with a Langevin equation, a stochastic first-order  differential equation that governs the time evolution of the system's observables \cite{Risken84,Strato2}.
Thanks to this link between FP equation and Langevin equation,  one can  determine the system's parameters once the drift and diffusion coefficients of the FP equation (first two Kramer-Moyals coefficients) are identified. 
The method has been applied successfully in many fields, including noisy electrical circuits, meteorological processes, traffic flow and physiological time series~\cite{Friedrich2011}; see also~\cite{Bodecker03} for an example of stochastic pitchfork bifurcation (in dissipative solitons). 
However, one fundamental limitation of this method lies in so-called \textit{finite-time effects}, that make inaccurate the direct evaluation of the Kramer-Moyals coefficients.
An alternative technique based on the adjoint Fokker-Planck equation was proposed by Honisch and Friedrich~\cite{Honisch11}, for computing these coefficients with a level of accuracy unaffected by finite-time effects.

In this paper we focus on Hopf bifurcations. Our main contribution is an extension of the aforementioned  adjoint-based SI method to the identification of the physical parameters  governing stochastic harmonic oscillators. Indeed, for harmonic oscillators, the above analysis can be pursued one step further from the Langevin equation back to the oscillator equation (\ref{eq:eta0}), and one can identify the actual physical parameters such as damping and stiffness.
We illustrate the method with the example of thermoacoustic systems, as our study is motivated by the practical relevance of identifying linear growth rates in the context of thermoacoustic instabilities in combustion chambers for gas turbines, aero- and rocket engines.
We are particularly interested in identifying nonlinear damping, determinant for stability, and leave aside stiffness nonlinearities.
We neglect multiplicative noise and focus on additive noise.
We assume that the system is made of a series of  harmonic oscillators, and we therefore proceed to perform SI independently for each oscillator by filtering time signals around the frequency of interest.
The paper is organised as follows. Section~\ref{sec:model} introduces  concepts that are useful to describe nonlinear harmonic oscillators dynamically and statistically, and that are necessary for system identification: Langevin equation, Fokker-Planck equation, and Kramer-Moyals (KM) coefficients.
Output-only system identification is the object of section~\ref{sec:SI}, which describes in detail how to compute finite-time KM coefficients, explains the limitations involved in extrapolation-based SI,
and  presents the more accurate adjoint-based SI.

\begin{figure}[] 
\psfrag{t}[t][]{$t$}
\psfrag{Peta}[t][]{\textcolor{red}{$P_\infty(\eta)$}}
\psfrag{PA}[t][]{$P_\infty(A)$}
\psfrag{etaA}[][]{\textcolor{red}{$\eta$}, $A$}
\centerline{
   \hspace{1cm}
   \begin{overpic}[height=6.5cm,tics=10]{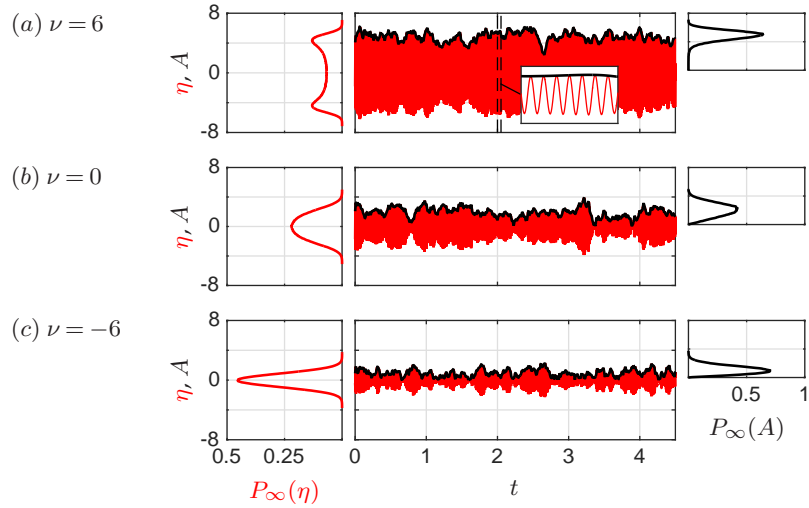}   
      \put(-25,73){$(a)$~$\nu=6$}
      \put(-25,49){$(b)$~$\nu=0$}
      \put(-25,25){$(c)$~$\nu=-6$}      
   \end{overpic} 
}
\caption{
Time signals of acoustic pressure modal amplitude  $\eta(t)$ and acoustic envelope $A(t)$, and  stationary PDFs $P_\infty(\eta)$ and $P_\infty(A)$ obtained from  simulations  of 
the stochastic Van der Pol oscillator (\ref{eq:eta3}) in three different regimes:
$(a)$~linearly unstable ($\nu=6$), $(b)$~marginally stable ($\nu=0$)
and $(c)$~linearly stable ($\nu=-6$).
Other parameters: $\kappa=2$, 
$\Gamma/4\omega_0^2=5$, 
$\omega_0/2\pi=150$~Hz.
} 
\label{fig:signals}
\end{figure}

\section{Stochastic oscillator model}
\label{sec:model}

\subsection{Dynamical description}

\begin{figure}[] 
\psfrag{A}[t][]{$A$}
\psfrag{AA}[t][]{$A$}
\psfrag{t}     [t][][1][-90]{$\,\, t$}
\psfrag{P(0)}  [][][1][-90]{$P(A,0) \,\,\,\,\,\,$}
\psfrag{P(end)}[][][1][-90]{$P_\infty(A) \,\,\,\,\,$}
\psfrag{D1}[][][1][-90]{$D^{(1)}(A)$}
\psfrag{VA}[][][1][-90]{$\mathcal{V}(A) \,\,\,\,\,\,\,$}
\psfrag{PA}[][][1][-90]{$\,\,\,\,\,\,P_\infty(A)$}
\centerline{
   \hspace{-0.6cm}
   \begin{overpic}[height=6.8cm,tics=10]{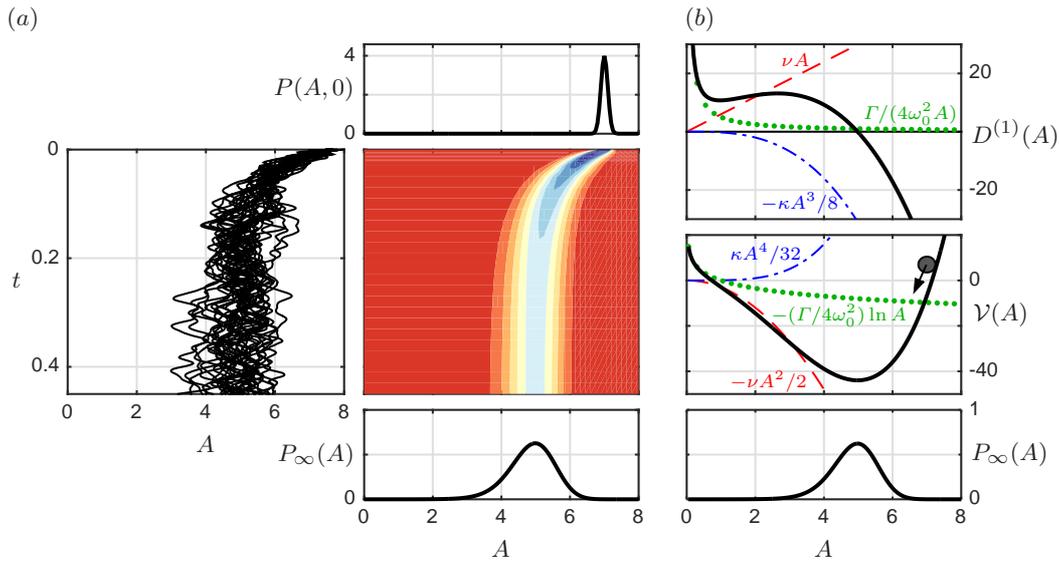} 
      \put( 0.5, 52){$(a)$}
      \put(66.6, 52){$(b)$}       
      \put(76,48){\textcolor{red}{\scriptsize $\nu A $}}
      \put(84,42.8){\textcolor[rgb]{0,0.7,0}{\scriptsize $\Gamma/(4\omega_0^2  A)$}}  
      \put(74,34){\textcolor{blue}{\scriptsize $-\kappa A^3/8$}}
      \put(71,29){\textcolor{blue}{\scriptsize$\kappa A^4/32$}}
      \put(75,23){\textcolor[rgb]{0,0.7,0}{\scriptsize$-(\Gamma/4\omega_0^2) \ln A$}}  
      \put(71,16.5){\textcolor{red}{\scriptsize $-\nu A^2/2 $}}   
   \end{overpic} 
}
\caption{
$(a)$~Left: acoustic amplitude $A(t)$ for 30 independent realisations of the stochastic VdP oscillator (\ref{eq:eta3}) starting from $\eta=7$, $\dot\eta=0$;
Right: time evolution of the PDF  $P(A,t)$  governed by the FP equation~(\ref{eq:FPE}). 
The PDF drifts and diffuses with time, finally converging to the stationary PDF $P_\infty(A)$.
$(b)$~Top: first KM coefficient $D^{(1)}(A)$, eq. (\ref{eq:D1D2});
Middle: acoustic potential $\mathcal{V}(A)$,  eq. (\ref{eq:potential});
Bottom: stationary PDF $P_\infty(A)$, eq. (\ref{eq:PDF_stat}).
Individual contributions to $D^{(1)}(A)$ and $\mathcal{V}(A)$ shown with dashed line (linear growth rate), dash-dotted line (nonlinearity) and dotted line (noise-induced term).
Parameters: $\nu=6$, $\kappa=2$, $\Gamma/4\omega_0^2=5$.
} 
\label{fig:FP}
\end{figure}

We consider the noise-driven harmonic oscillator
\begin{equation}
\ddot \eta  + \omega_{0}^2 \eta =  f(\eta,\dot\eta) + \xi(t),
\label{eq:eta1}
\end{equation}
where  $\xi(t)$ is a white 
Gaussian noise  of intensity $\Gamma$, i.e. $\langle \xi\xi_\tau\rangle = \Gamma \delta(\tau)$.
Equation (\ref{eq:eta1}) describes the dynamics of one of possibly many oscillators, in the absence of strong coupling.
In this case, the state $\eta(t)$ can be isolated by band-pass filtering the signal, provided $\omega_0$ is well separated from the natural frequency of all other oscillators.
Since $\eta(t)$ is quasi-harmonic, it  can be written
$\eta(t)=A(t) \cos(\omega_0 t + \varphi(t))$ 
where the the envelope  $A(t)$ and phase $\varphi(t)$ are varying slowly with respect to the  period $2\pi/\omega_0$ (figure~\ref{fig:signals}).
Expanding $f$ with a Taylor series,
\begin{equation}
f(\eta, \dot\eta) = \sum_n \sum_m  \gamma_{n,m} \eta^n \dot\eta^m,
\label{eq:fff}
\end{equation}
substituting into the expressions 
$A(t) = \sqrt{\eta^2+(\dot\eta/\omega_0)^2}$ and
$\varphi(t) = -\mbox{atan}(\dot\eta/\omega_0 \eta)-\omega_0 t$,
and performing  deterministic and stochastic averaging   \cite{Strato2,Roberts86} yields a set of 
stochastic differential equations (Langevin equations),
\begin{align}
\displaystyle
\dot A
&= 
\frac{\gamma_{0,1}}{2} A
+\left( \frac{\gamma_{2,1}}{8}+\frac{3\omega_0^2 \gamma_{0,3}}{8} \right) A^3
+\frac{\Gamma}{4\omega_0^2 A}
+ \zeta(t) + O(A^5),
\label{eq:Langevin-A-general}
\\
\displaystyle 
\dot \varphi
&= 
-\frac{\gamma_{1,0}}{2\omega_0} 
-\left( \frac{\omega_0 \gamma_{1,2}}{8}+\frac{3\gamma_{3,0}}{8 \omega_0} \right) A^2
+ \frac{1}{A} \chi(t) + O(A^4),
\label{eq:Langevin-phi-general}
\end{align}
where $\zeta(t)$ and $\chi(t)$ are two independent white Gaussian noises of intensity $\Gamma/2\omega_0^2$, i.e. $\langle \zeta\zeta_\tau\rangle = \langle \chi\chi_\tau\rangle = (\Gamma/2\omega_0^2) \delta(\tau)$.
Equations (\ref{eq:Langevin-A-general})-(\ref{eq:Langevin-phi-general}) are valid up to $A^4$ for any $f$.
Focusing on the equation for the envelope $A(t)$  
that is decoupled from that for the phase $\varphi(t)$, one can rewrite
 \begin{align}
\displaystyle 
\dot A
&= \nu A - \frac{\kappa}{8} A^3 + \frac{\Gamma}{4 \omega_0^2 A} + \zeta(t)
= -\frac{\mathrm{d} \mathcal{V}}{\mathrm{d} A} + \zeta(t).
\label{eq:Langevin-A}
\end{align}
Here $\nu$ is the linear  growth rate of the system, whose sign determines the oscillator's linear stability.
Saturating nonlinear effects come into play via $\kappa>0$.
The potential  
\begin{equation}
\mathcal{V}(A)
= - \frac{\nu}{2} A^2  + \frac{\kappa}{32} A^4 - \frac{\Gamma}{4 \omega_0^2} \ln(A)
\label{eq:potential}
\end{equation}
thus governs the dynamics of $A(t)$ 
and can be decomposed as follows (figure~\ref{fig:FP}$b$): the envelope $A$ is attracted  (resp. repelled) by the stable (resp. unstable) low-amplitude  equilibrium  when the growth rate $\nu$ is negative (resp. positive); the nonlinearity $\kappa$ prevents $A$ from diverging to infinity; and the deterministic noise contribution $\Gamma$ prevents $A$ from vanishing.
Note how  the additive noise $\xi(t)$ that acts as a stochastic forcing for $\eta(t)$ in (\ref{eq:eta1}) has a twofold effect on $A(t)$:  deterministic contribution proportional to $\Gamma$ and stochastic forcing $\zeta(t)$.
Note also that $A^2=\eta^2+(\dot\eta/\omega_0)^2$ is generally proportional to the sum of a potential energy and a kinetic energy.

In this paper we  illustrate  system identification  with the amplitude equation (\ref{eq:Langevin-A}) and one of its possible associated oscillators, namely the stochastic Van der Pol (VdP) oscillator
\begin{equation}
\ddot \eta - 2\nu \dot \eta + \omega_0^2 \eta 
=  -\kappa \eta^2 \dot \eta + \xi(t),
\label{eq:eta3}
\end{equation}
which corresponds to the cubic nonlinearity 
$f=\mathrm{d}(2\nu\eta-\kappa\eta^3/3)/\mathrm{d}{t}=
2\nu\dot\eta-\kappa\eta^2\dot\eta$ at $\omega_0$ (i.e. $\gamma_{2,1}=-\kappa$), and which is characterised in the purely deterministic case by a supercritical Hopf bifurcation occurring at $\nu=0$.
The method would apply as is, however,  if higher-order terms needed to describe $f(\eta,\dot \eta)$ were included (e.g. quintic terms for  subcritical bifurcations \cite{Noiray16,Gopal2016}).
Figure~\ref{fig:signals} shows typical signals $\eta(t)$ and $A(t)$ obtained by simulating (\ref{eq:eta3}) with different linear growth rates. 

 In the specific case of thermoacoustic systems in combustion chambers, equation (\ref{eq:eta1}) faithfully reproduces the dynamics of pressure oscillations associated with one thermoacoustic mode~\cite{NoiraySchu13,Noiray16,NoirayDenisov16}.
The pressure field is projected with a Galerkin method onto a basis of space-dependent acoustic eigenmodes with time-dependent coefficients $\eta(t)$ \cite{CulickAGARD,lieuwen_book_2012},
and band-pas filtering around the mode's frequency  isolates its contribution and yields a quasi-harmonic signal \cite{Culick76a,Lieuwen03}.
Heat release rate fluctuations from the flame are the sum of 
coherent fluctuations induced by the acoustic field,
and stochastic fluctuations induced by turbulent flow perturbations whose  spatial correlations are much smaller than the acoustic wavelength and which can be modeled by $\xi$. (Colored noise could readily be taken into account, see e.g. \cite{boncioliniboujonoiray16}.)
The linear thermoacoustic growth rate $\nu=(\beta-\alpha)/2$ of the mode of interest is the result of the competition between the acoustic damping $\alpha>0$ and the linear contribution of the flame $\beta>0$  or  $<0$, while $\kappa>0$ comes from the flame nonlinearities.

\subsection{Statistical description}

In a purely deterministic case, the amplitude equation (\ref{eq:Langevin-A}) is a Stuart-Landau equation,
\begin{equation}
\displaystyle
\dot A = \nu A - \frac{\kappa}{8} A^3,
\end{equation}
whose long-time solution is either the fixed point $A_{det}=0$ when the system is linearly stable ($\nu<0$), or the limit-cycle of amplitude $A_{det}=\sqrt{8\nu/\kappa}$ when the system is linearly unstable ($\nu>0$).

In the stochastic case, the envelope is fluctuating in time and it is convenient to adopt a statistical description of the system. Examples of  probability density functions (PDFs) in the stationary regime are shown in figure~\ref{fig:signals}. 
$P_\infty(\eta)$ is symmetric about $\eta=0$, and  shifts from a unimodal distribution (maximum for $\eta=0$) to a bimodal distribution (maxima in $|\eta|>0$) as  $\nu$ increases and the system becomes linearly unstable \cite{Lieuwen03}.
Accordingly, the most probable amplitude $A_{max}$ moves toward larger values and an inflection point appears between $A=0$ and $A_{max}$.
More generally, the  evolution of the PDF $P(A,t)$ is governed by the Fokker-Planck  equation associated with (\ref{eq:Langevin-A}):
\begin{align}
\frac{\partial }{\partial t} P(A,t) 
&=                         -\frac{\partial   }{\partial A}   \left( D^{(1)} P(A,t) \right) 
+                           \frac{\partial^2 }{\partial A^2} \left( D^{(2)} P(A,t) \right),
\label{eq:FPE}
\\
D^{(1)}(A) &= \nu A - \frac{\kappa}{8} A^3 + \frac{\Gamma}{4 \omega_0^2 A},
\quad 
D^{(2)}(A) = \frac{\Gamma}{4\omega_0^2}.
\label{eq:D1D2}
\end{align}
The FP equation is a particular type of convection-diffusion equation (figure~\ref{fig:FP}$a$), whose drift and diffusion coefficients $D^{(n)}$, $n=1,2$, are also called the first two Kramers-Moyal  coefficients \cite{Risken84,Strato2}.
With our specific choice of system (additive noise only has been considered), $D^{(2)}$ is independent of $A$.
The stationary PDF of $A$ is  the long-time solution of (\ref{eq:FPE}), and reads explicitly here
\begin{equation}
P_\infty(A) = \lim_{t \rightarrow \infty} P(A,t) 
= \mathcal{N} \exp \left( \frac{1}{D^{(2)}} \int_0^A D^{(1)}(A') \,\mathrm{d}A' \right)
= \mathcal{N} \exp \left( -\frac{4\omega_0^2}{\Gamma} \mathcal{V(A)} \right),
\label{eq:PDF_stat}
\end{equation}
with $\mathcal{N}$ a normalisation coefficient such that $\int_0^\infty P_\infty(A) \,\mathrm{d}A = 1$.
Therefore, $P_\infty(A)$ is directly determined by the KM coefficients $D^{(1)}$ and $D^{(2)}$, or equivalently by the potential $\mathcal{V(A)}$ and the noise intensity $\Gamma$.
Note that  $A_{max}$ corresponds to the zero of $D^{(1)}(A)$ and the minimum of  $\mathcal{V(A)}$, and is in general different both from the time-averaged amplitude and from the deterministic amplitude.

\section{System identification with the KM coefficients}
\label{sec:SI}

In the context of output-only system identification,
we are interested in finding 
the system's parameters without any actuation devices (that are typically employed to study impulse response of harmonic response),
but based instead on an acoustic signal measured in uncontrolled conditions. 
We take advantage of the inherent external forcing (coming from intense turbulence in the case of thermoacoustics in combustors), which drives the system away from its purely deterministic limit cycle $A(t)=A_{det}$ and allows one to retrieve precious information in a range of amplitudes $A$.

The expression (\ref{eq:PDF_stat}) of the stationary PDF depends explicitly  on the system's parameters (here $\nu$,  $\kappa$ and  $\Gamma$), which can therefore be  identified by fitting the analytical expression to the measured PDF. 
(More precisely, $P_\infty$ depends on the two ratios $\nu/\Gamma$ and  $\kappa/\Gamma$; identifying unambiguously the three parameters requires using one additional method, using for instance the power spectral density of $\eta$ in the linearly stable regime, or the power spectral density of fluctuations of $A$ in the unstable regime.)
Noiray and Schuermans~\cite{NoiraySchu13} proposed another system identification method based on estimating the Kramers-Moyal coefficients and fitting the analytical expressions (\ref{eq:D1D2}) and applied it to data from a gas turbine combustor.
Noiray and Denisov~\cite{NoirayDenisov16} validated this SI method with a lab-scale combustor, comparing transient regimes calculated numerically with the FP equation (solved using the identified parameters) to transient regimes
measured in series of ON$\rightarrow$OFF and OFF$\rightarrow$ON control switching.
The principle of the method is recalled in section~\ref{sec:SI}\ref{sec:extrap}.
In practical combustors, one needs to band-pass filter the acoustic signal prior to SI in order to isolate the dynamics of the thermoacoustic mode of interest from the dynamics of other modes. 
(Recall that we use a single-mode approximation. One might consider a more complex description of the system, with several modes and therefore more parameters to identify. It should be underlined that increasing the number of parameters might affect negatively the reliability of the identification, thus it is preferable to limit the macroscopic model to its essential ingredients.)

As explained in section~\ref{sec:SI}\ref{sec:extrap}, this band-pass filtering  hinders the application of this SI method.
In this paper, we present a new SI method that uses the KM coefficients too, but is based on a different approach (section~\ref{sec:SI}\ref{sec:exact}) and helps circumventing the fundamental limitation of the aforementioned method.
Before proceeding, we  first detail in section~\ref{sec:SI}\ref{sec:finiteKM}  how the KM coefficients can be estimated from measurements.

\subsection{Finite-time KM coefficients}
\label{sec:finiteKM}

\begin{figure}[] 
\psfrag{t}[t][]{$t$ [s]}
\psfrag{a}[t][]{$a$}
\psfrag{A(t+tau)}[t][]    {$A(t+\tau)$}
\psfrag{A(t)}[r][][1][-90]{$A(t)$}
\psfrag{A}[r][][1][-90]{$A$}
\psfrag{PPP}[r][][1][-90]{}
\psfrag{M1}[r][][1][-90]{}
\psfrag{M2}[r][][1][-90]{}
\centerline{
   \hspace{0.5cm}
   \begin{overpic}[height=5cm,tics=10]{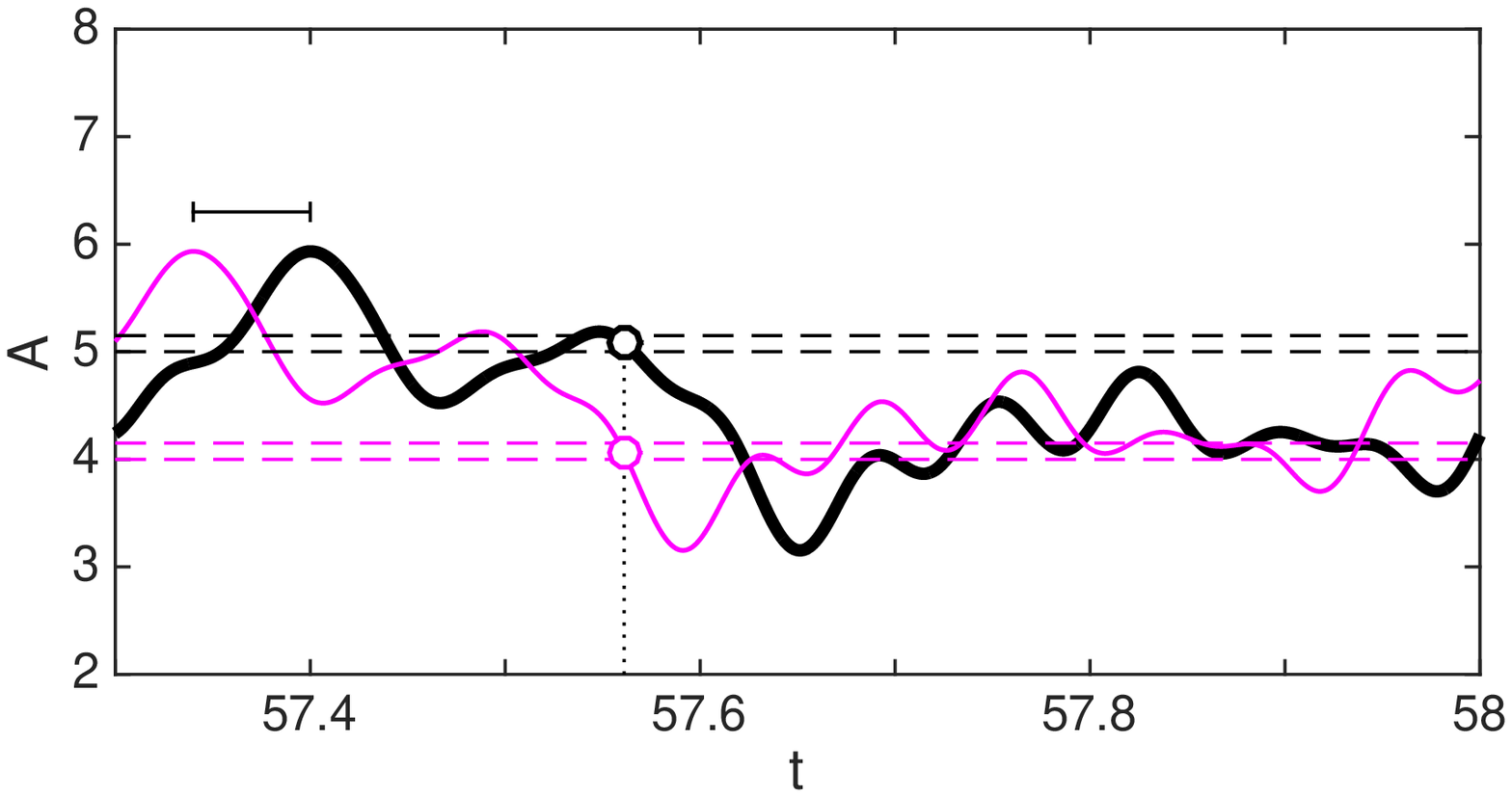}    
      \put( -2,49){$(a)$}
      \put(16,42){$\tau$}     
      \put(45,33){\small $A(t)=5$}
      \put(19,19){\small \textcolor[rgb]{1,0,1}{$A(t+\tau)=4$}}
   \end{overpic} 
}
\vspace{1cm}
\centerline{
   \hspace{1.1cm}
   \begin{overpic}[height=4.5cm,tics=10]{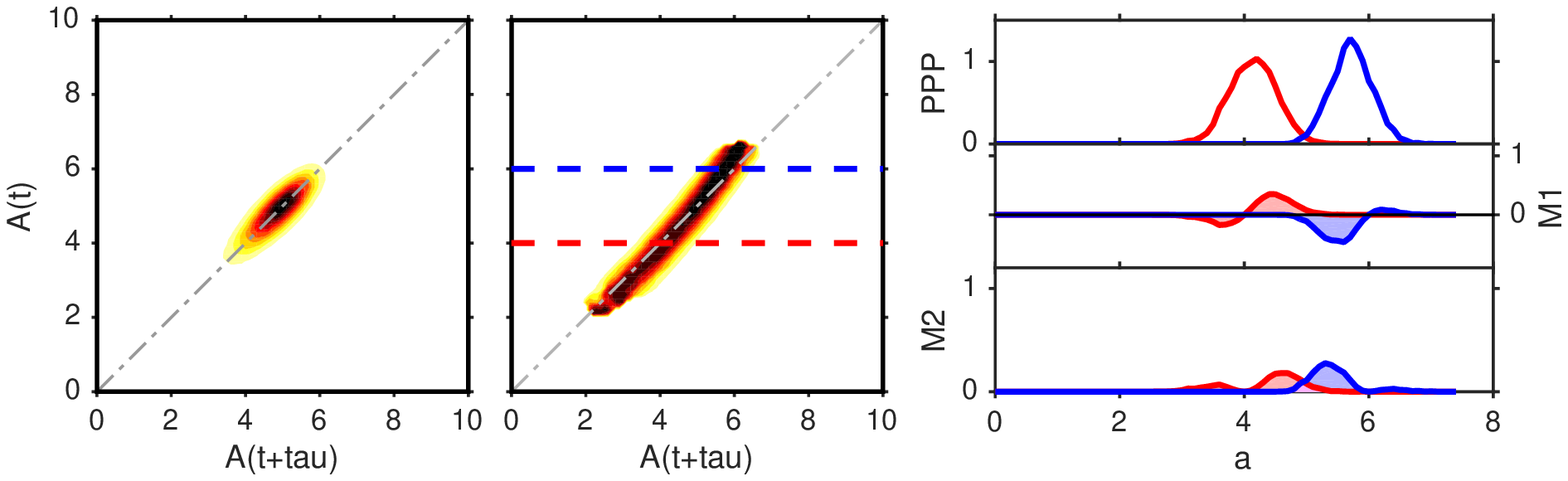}
      \put( 6,31){$(b)$ $P(a(t+\tau),A(t))$}
      \put(33,31){$(c)$ $P(a,t+\tau|A,t)$}
      \put(64,31){$(d)$}
      \put(7,27){$\tau=0.02$~s}
      \put(49,17.7){\small \textcolor{blue}{$A=6$}}     
      \put(49,13.0){\small \textcolor{red}{$A=4$}}
      \put(71.0,23){\small \textcolor{red}{$A=4$}}
      \put(88.7,23){\small \textcolor{blue}{$A=6$}}
      \put(64,26.9){\small $P(a,t+\tau|A,t)$}
      \put(64,18.9){\small $(a-A)   P(a,t+\tau|A,t)$}
      \put(64,10.9){\small $(a-A)^2 P(a,t+\tau|A,t)$}       
   \end{overpic} 
}
\vspace{-0.25cm}
\centerline{   
   \hspace{1.1cm}
   \begin{overpic}[height=4.5cm,tics=10]{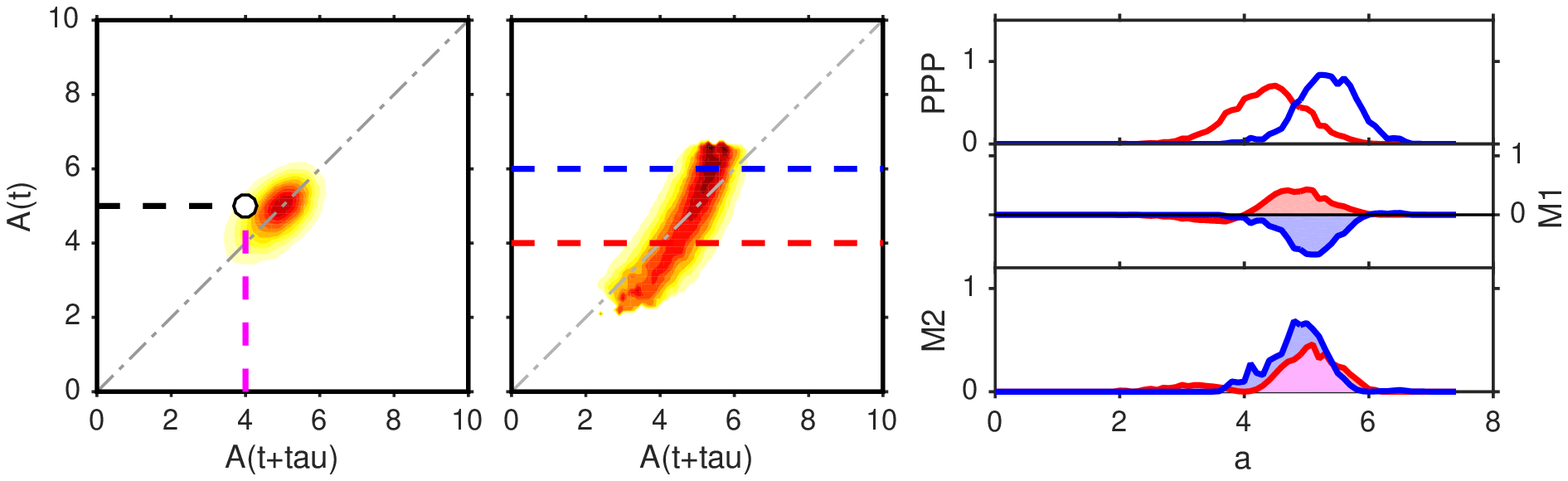}
      \put(7,27){$\tau=0.06$~s}
      \put(6.5,18){\small $A(t)=5$}
      \put(16,7){\small \textcolor[rgb]{1,0,1}{$A(t+\tau)=4$}}
   \end{overpic} 
}
\vspace{-0.25cm}
\centerline{
   \hspace{1.1cm}
   \begin{overpic}[height=4.5cm,tics=10]{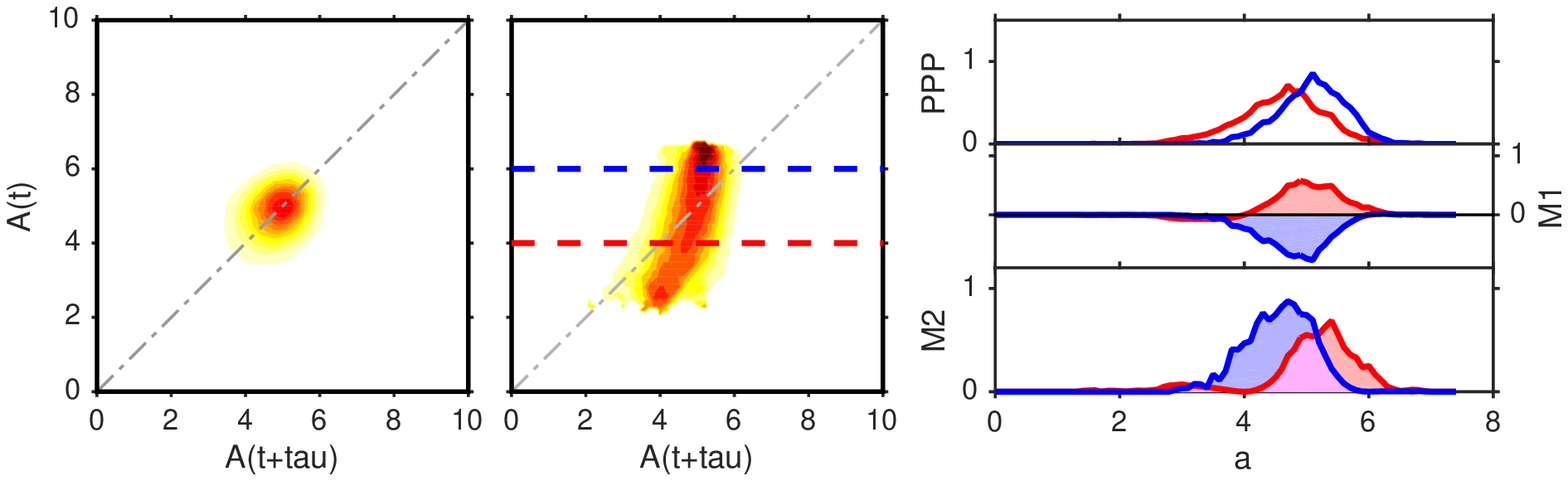}
      \put(7,27){$\tau=0.12$~s}
   \end{overpic} 
}
\caption{
Estimation of finite-time KM coefficients from time series.
$(a)$~Acoustic envelope signal (thick line) and time-shifted signal (thin line, shift $\tau=0.06$~s) used to compute the joint probability $P(a(t+\tau),A(t))$. For instance, the vertical dashed line shows one occurrence of the event $\{A(t)=5$, $A(t+\tau)=4\}$ that contributes to the joint probability  $P(a=4,A=5)$ indicated by the circle in panel $b$.
$(b)$~Joint probability $P(a(t+\tau),A(t))$ (not shown where $P_\infty$ is smaller than $1\%$ of its maximum).
$(c)$~Conditional probability $P(a,t+\tau|A,t)$.
$(d)$~One-dimensional cuts of the conditional probability 
at $A=4$ and $A=6$, and integrands $(a-A)^n P(a,t+\tau|A,t)$  
used to estimate the finite-time KM coefficients (\ref{eq:D}).
Parameters: $\nu=6$, $\kappa=2$, $\Gamma/4\omega_0^2=5$, $\omega_0/2\pi=150$~Hz, $T=1000$~s, $\Delta f=60$~Hz.
} 
\label{fig:moments}
\end{figure}

The Kramers-Moyal coefficients $D^{(n)}$, $n=1,2$, in the FP equation (\ref{eq:FPE}) can be computed from a time signal $A(t)$ as 
\begin{eqnarray}
D^{(n)}(A) = \lim_{\tau\rightarrow 0} D^{(n)}_\tau(A),
\qquad
D^{(n)}_\tau(A) = \frac{1}{n!\tau} \int_{0}^{\infty} (a-A)^n P(a,t+\tau|A,t) \,\mathrm{d}a, 
\label{eq:D}
\end{eqnarray}
i.e. as the short-time limit of  the \textit{finite-time} coefficients $D^{(n)}_\tau(A)$, which are related to the moments of the conditional PDF $P(a,t+\tau|A,t)$  describing   the probability of the amplitude being $a$ at time $t+\tau$ knowing that it is $A$ at time $t$~\cite{Risken84,Strato2}.
Finite-time KM coefficients $D^{(n)}_\tau$ are readily calculated for a given finite time shift $\tau>0$ by processing a time signal measured in the stationary regime, as illustrated in figure~\ref{fig:moments}: 
\begin{itemize}
\item 
The envelope $A(t)$ is calculated 
(using for instance the Hilbert transform~\cite{Feldman11}) from the band-pass filtered acoustic pressure signal $\eta(t)$;
\item 
Data binning of the envelope $A(t)$ and time-shifted envelope $A(t+\tau)$ (figure~\ref{fig:moments}$a$) yields the joint PDF $P(a(t+\tau),A(t))$ (figure~\ref{fig:moments}$b$); 
\item
The conditional PDF is then deduced from
$P(a,t+\tau|A,t) = P(a(t+\tau),A(t)) / P_\infty(A(t))$ (figure~\ref{fig:moments}$c$);
\item
Finite-time KM coefficients are finally obtained by computing moments of the conditional PDF according to (\ref{eq:D}) (figure~\ref{fig:moments}$d$).
\end{itemize}

At this point, it is worth commenting several features of figure~\ref{fig:moments}. (i)~For any value of $\tau$, the joint probability is symmetric about the diagonal 
$A(t)=A(t+\tau)$ and is  maximum for $A=A_{max}$, as expected in the stationary regime.
(ii)~The conditional probability is tilted with respect to  the diagonal around the point  $A=A_{max}$; this means that if at time $t$ there is an excursion $A(t) > A_{max}$  then it is  likely that $A$ will oscillate back to a lower value by the time $t+\tau$, and vice-versa.
(iii)~In the limit of small $\tau$ values, the joint probability tends to  $P(A) \delta(a-A)$, the conditional probability tends to $\delta(a-A)$, and the moments therefore necessarily tend to $\int_{0}^{\infty} (a-A)^n  \delta(a-A) \,\mathrm{d}a = 0.$
However, it is the linear rate at which the moments tend to 0 that defines the KM coefficients (\ref{eq:D}).
(iv)~In the limit of large $\tau$ values, any correlation between $A(t)$ and $A(t+\tau)$ is lost: 
the conditional probability tends to $P(a) \times P(A)$, the joint probability becomes tends to $P(a) \,\forall A$, and the KM coefficients become independent of $A$.

\begin{figure}[] 
\psfrag{tau}[t][]{$\tau$~[s]}
\psfrag{D1}[r][][1][-90]{$D^{(1)}_\tau$}
\psfrag{f}[t][]{$f$~[Hz]}
\psfrag{PSD}[r][][1][-90]{$\mathcal{S}_{\eta\eta}$}
\centerline{ 
   \hspace{0.3cm}
   \begin{overpic}[height=5.1cm,tics=10]{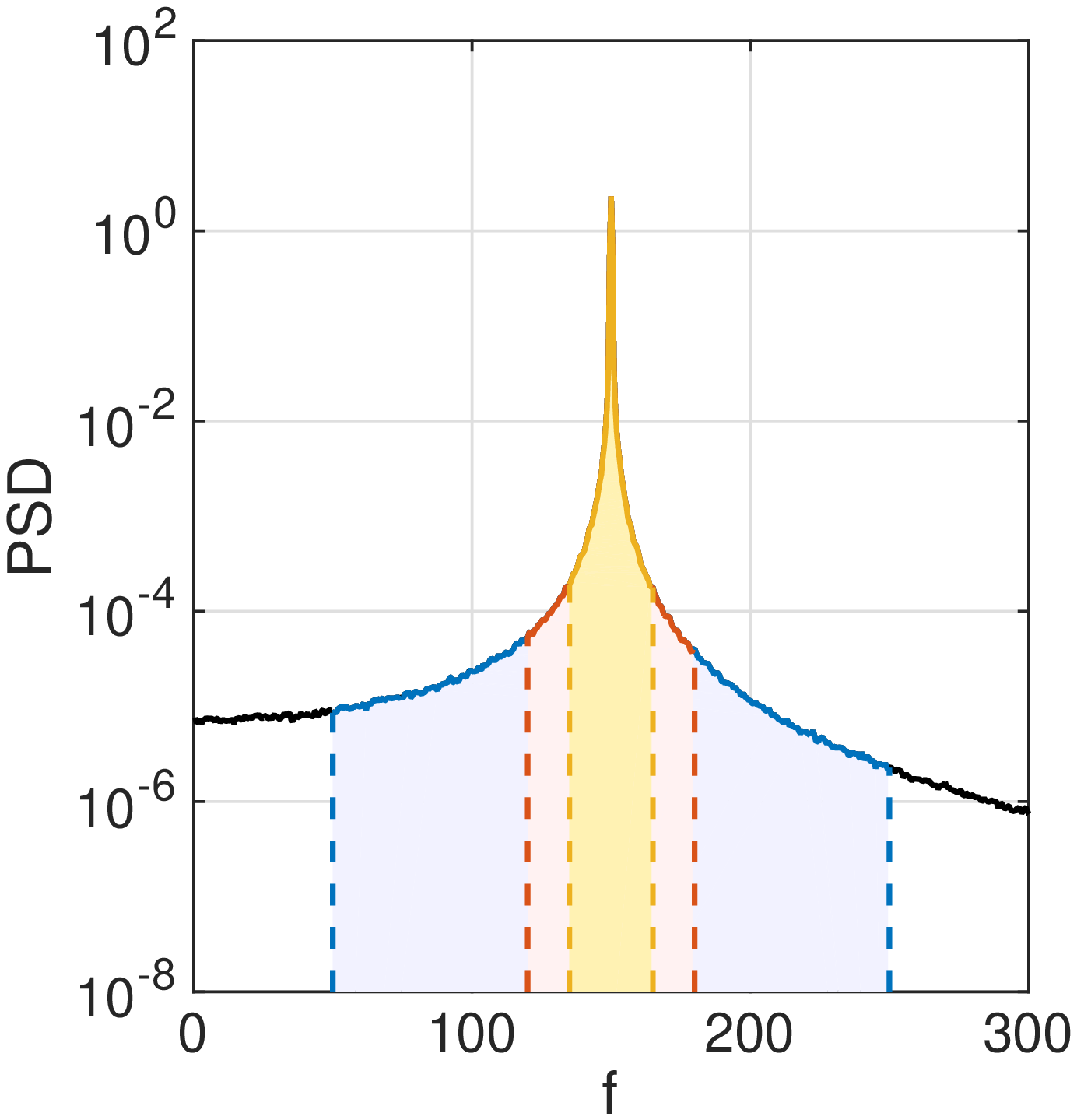}    
      \put( -8,92.5){$(a)$}   
   \end{overpic}    
   \hspace{0.7cm}
   \begin{overpic}[height=5.cm,tics=10]{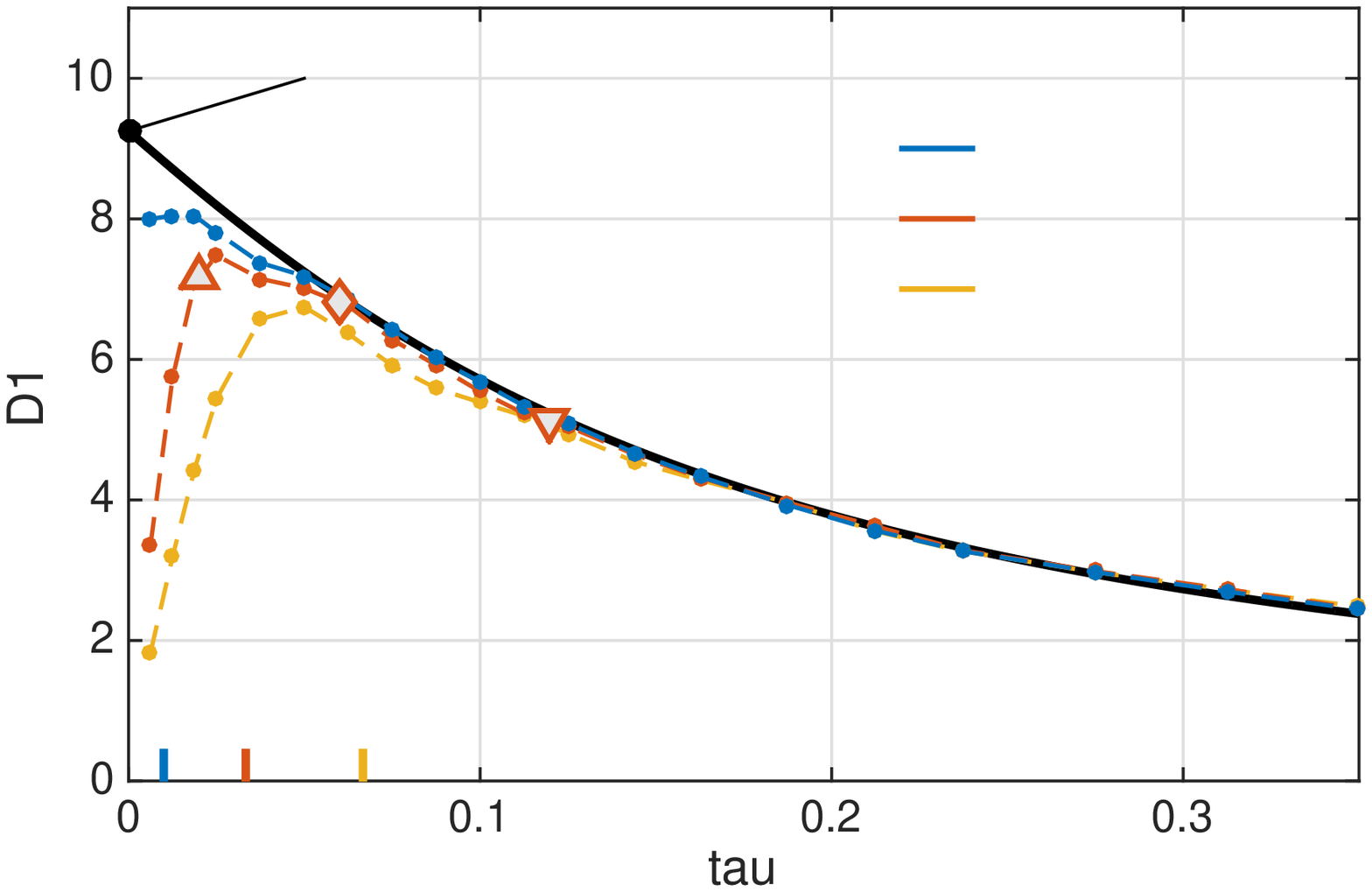} 
      \put( -6,62){$(b)$}     
      \put( 23,59){$\displaystyle D^{(1)} = \lim_{\tau\rightarrow 0} D^{(1)}_\tau$}        
      \put( 72, 54.){$\Delta f=200$ Hz}     
      \put( 72, 48.75){$\Delta f=60$ Hz}     
      \put( 72, 43.5){$\Delta f=30$ Hz}     
      \put( 15,12.5){$2/\Delta f$}     
   \end{overpic}      
}
\caption{
Illustration of finite-time effects: the  finite-time KM coefficient $D^{(1)}_\tau$ calculated from the envelope $A(t)$ of a time signal $\eta(t)$ filtered with different bandwidths 
$\Delta f$ (panel $a$) deviates from its theoretical value for small time shifts $\tau \lesssim 1/\Delta f$ (panel $b$).
Solid line: theoretical value obtained from the AFP equation.
Dot at $\tau=0$: exact KM coefficient $D^{(1)}$.
Symbols at $\tau=0.02$, $0.06$ and $0.12$ correspond to figure~\ref{fig:moments}.
Thicker ticks: $\tau \propto 1/\Delta f$. 
Parameters: $\nu=6$, $\kappa=2$, $\Gamma/4\omega_0^2=5$, $\omega_0/2\pi=150$~Hz, $T=500$~s, $A=4$, $\Delta f=30$, $60$ and $200$~Hz.
} 
\label{fig:D}
\end{figure}

In practice, computing the limit in (\ref{eq:D}) for infinitesimally small time shifts $\tau \rightarrow 0$ might be a delicate task.
This is illustrated in figure~\ref{fig:D}, which shows a typical example of how $D_\tau^{(1)}$ varies with $\tau$ for a given value of $A$.
It appears indeed that, when estimated from time signals, finite-time KM coefficients deviate from their theoretical value as $\tau$ becomes small.
This is caused by one or several \textit{finite-time effects}:
the data acquisition sampling rate is finite;
the actual noise is not strictly $\delta$-correlated and the Markov property necessary to make use of (\ref{eq:D}) does not hold; band-pass filtering removes high-frequency (i.e. small-time) information from the signal.
In combustion chambers, filtering is generally the dominant effect due to the need to isolate the thermoacoustic mode of interest when one intends to do SI using a single-mode description.

\begin{figure}[] 
\psfrag{A=4}[][]{}
\psfrag{RAW}[][]{}
\psfrag{A}[t][]{$A$}
\psfrag{D1}[r][][1][-90]{$\widehat D^{(1)}_\tau(A)$}
\psfrag{DDD}[r][][1][-90]{$\widehat D^{(1)}(A)$}
\psfrag{tau}[t][]{$\tau$~[s]}
\centerline{
   \begin{overpic}[height=6cm,tics=10]{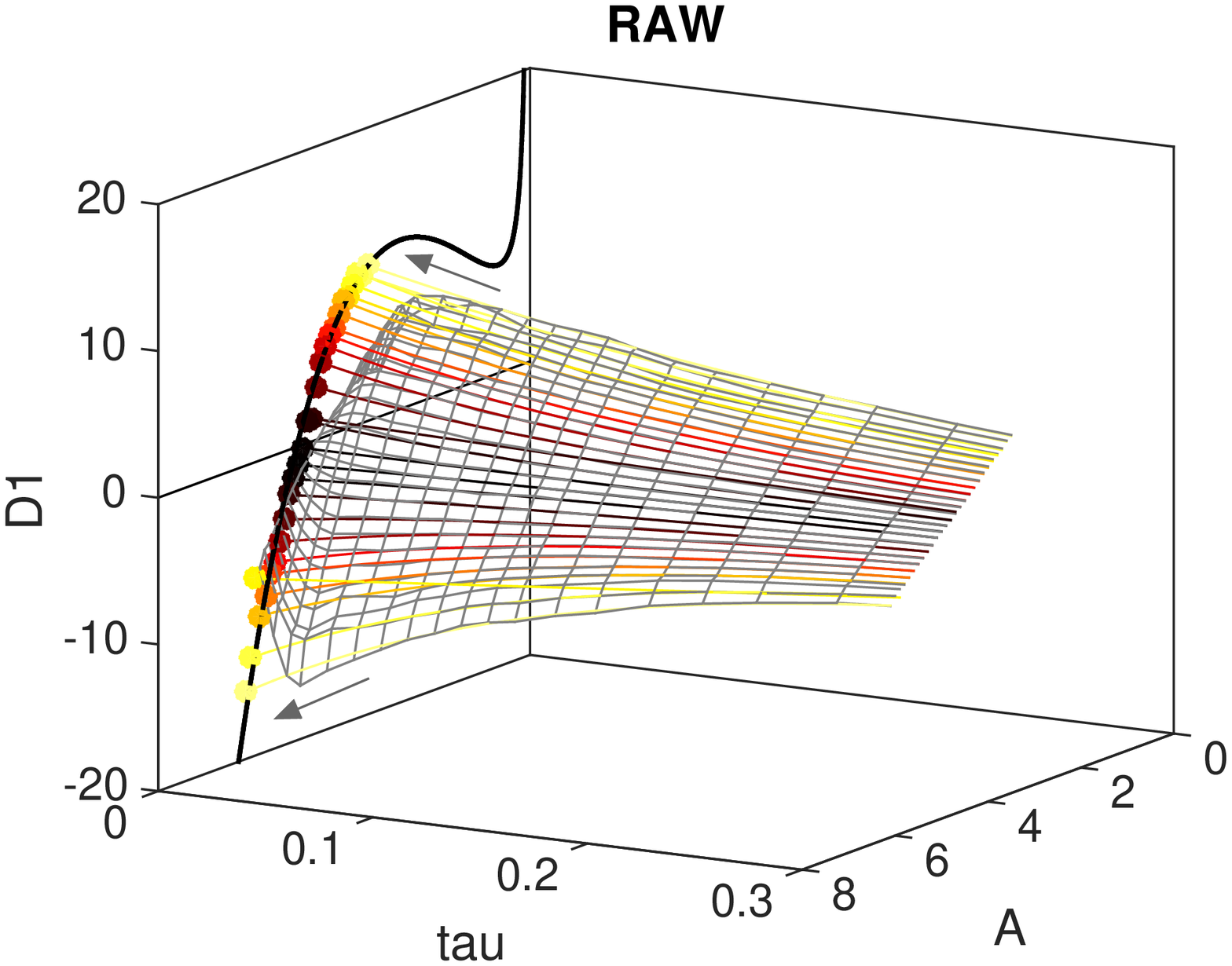} 
      \put(-10,68){$(a)$} 
      \put( 20,56){ \rotatebox{30}{$\widehat D^{(1)}$} }  
   \end{overpic}  
}
\vspace{1cm}
\centerline{
   \hspace{1.cm}
   \begin{overpic}[height=4.9cm,tics=10]{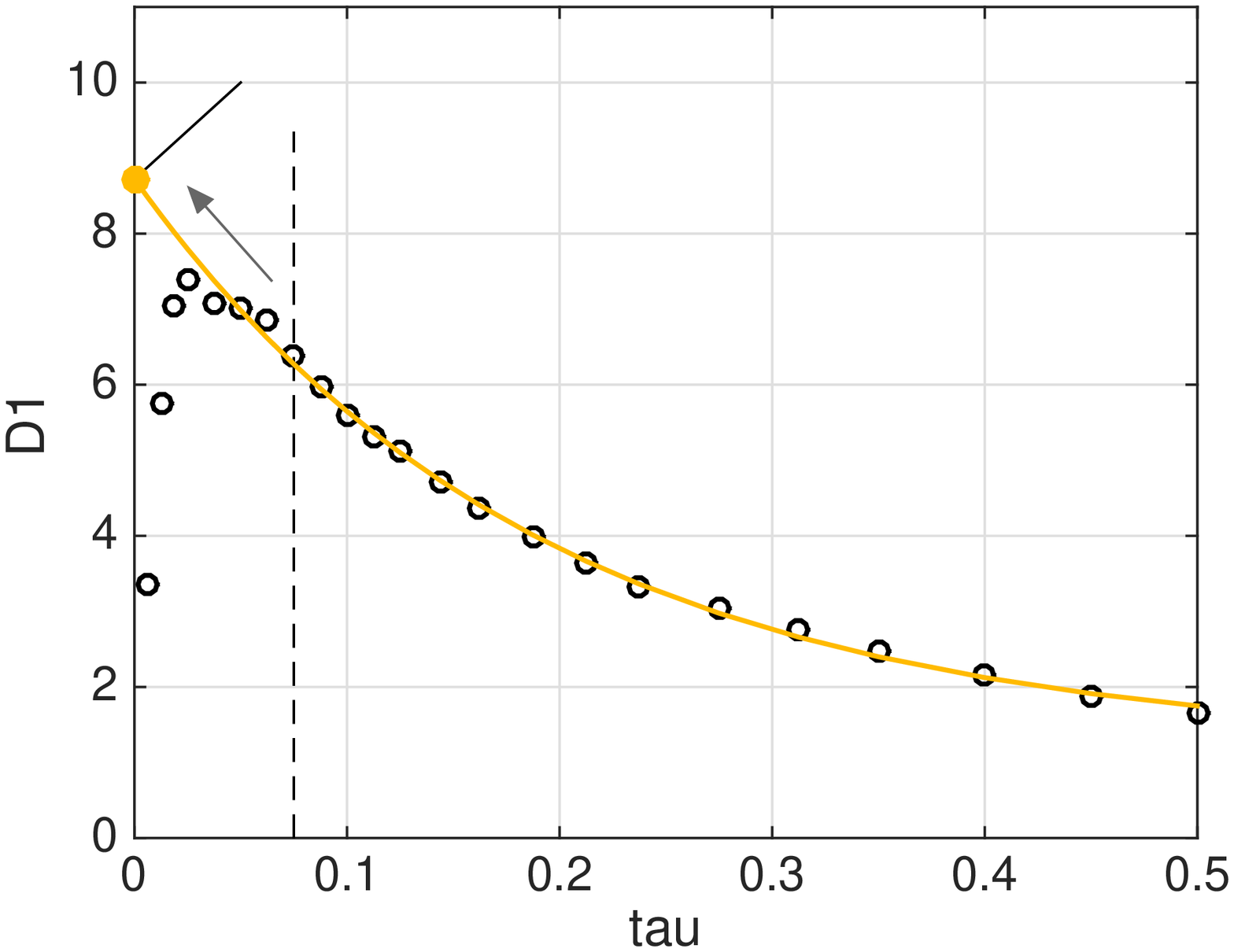} 
      \put(-10,73){$(b)$}
      \put( 21,70){$\displaystyle \widehat  D^{(1)} = \lim_{\tau\rightarrow 0} \widehat  D^{(1)}_\tau$}  
   \end{overpic} 
   \hspace{1cm}
   \begin{overpic}[height=4.9cm,tics=10]{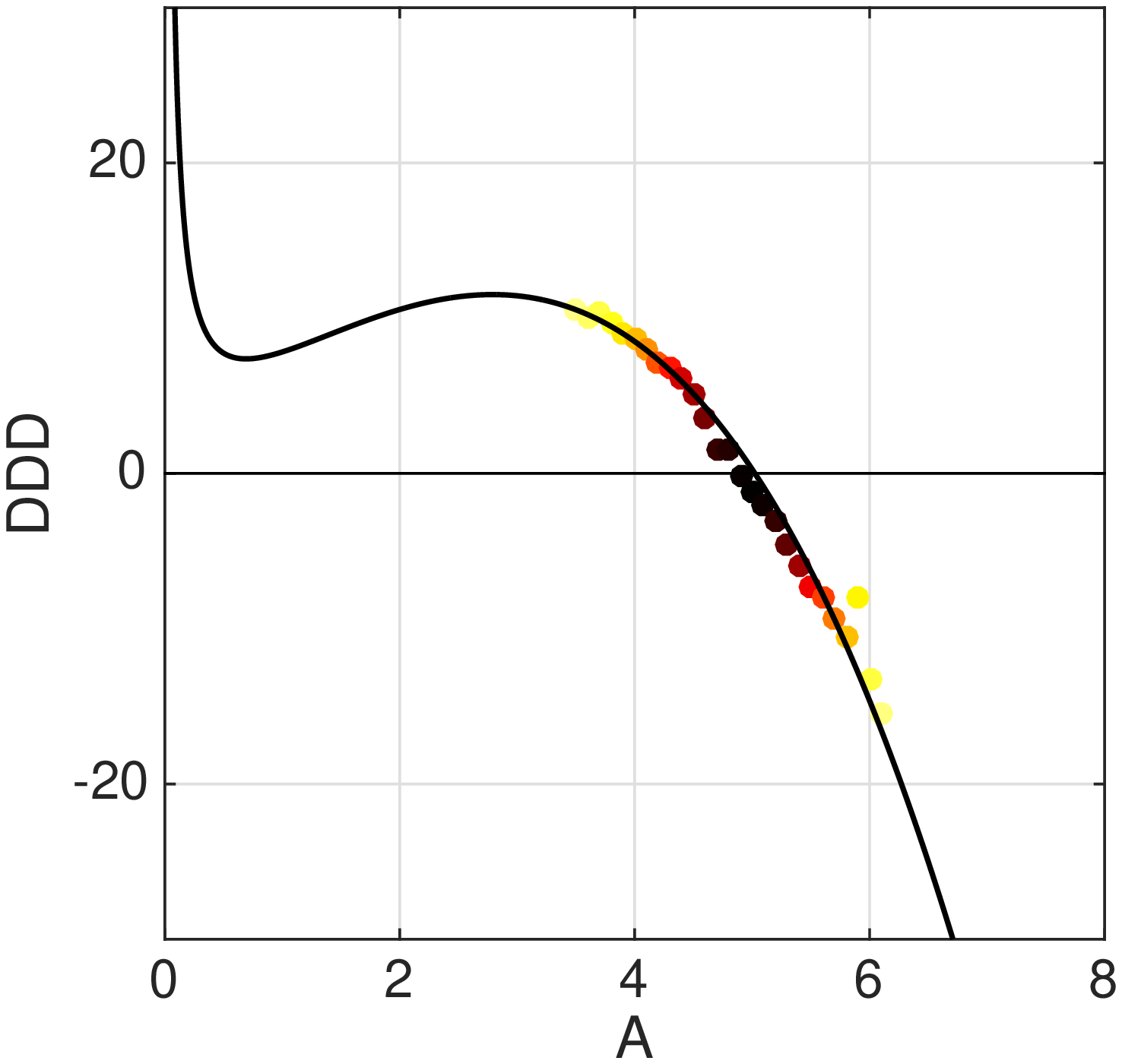} 
      \put(-10,91){$(c)$}
   \end{overpic}
}
\caption{
System identification based on extrapolation.
Estimated finite-time KM coefficients $\widehat D^{(1)}_\tau(A)$ are interpolated over a range of time shifts $\tau$ where finite-time effects are negligible, and extrapolated to $\tau \rightarrow 0$ (see e.g. panel $(b)$ for $A=4$).
The extrapolation $\displaystyle \widehat D^{(1)} = \lim_{\tau\rightarrow 0} \widehat D^{(1)}_\tau$ is repeated independently for each amplitude $A$ (panel $(a)$).
Finally, extrapolated values  are fitted with the analytical expression (\ref{eq:D1D2}), allowing the identification of $\{\nu,\kappa,\Gamma\}$ (panel $(c)$). Parameters: $\nu=6$, $\kappa=2$, $\Gamma/4\omega_0^2=5$,  $\omega_0/2\pi=150$~Hz, $T=500$~s, $\Delta f=60$~Hz.
} 
\label{fig:extrap_3D}
\end{figure}

\subsection{Extrapolation of finite-time KM coefficients to $\tau=0$}
\label{sec:extrap}

In order to avoid finite-time effects, finite-time KM coefficients  $D^{(n)}_\tau(A)$ can  be calculated for large enough time shifts  $\tau$, and the exact KM coefficients $D^{(n)}(A)$ can be estimated by interpolating the data and carefully extrapolating to $\tau=0$, as shown in figure~\ref{fig:extrap_3D}$a,b$.
Analytical expressions that can be derived for $D^{(n)}_\tau(A)$ in simple cases such as the Ornstein-Uhlenbeck process \cite{Risken84,Honisch11} suggest using exponential functions of $\tau$ for the interpolation; we are not aware of analytical expressions in more complex cases.
Alternatively, one could use the moments 
$n!\tau D^{(n)}_\tau(A) =  \int_{0}^{\infty} (a-A)^n P(a,t+\tau|A,t) \,\mathrm{d}a$ to estimate the KM coefficients $D^{(n)}(A)$.
For more details, the reader is referred to \cite{Friedrich97PRL,Friedrich97PhysD} and \cite{Friedrich2011}, where the issue is discussed at length with both theoretical aspects and many application examples.

Repeating for different amplitudes $A$ and fitting to the extrapolated values the analytical expressions (\ref{eq:D1D2}) of $D^{(n)}(A)$, allows the identification of the parameters $\nu$, $\kappa$ and $\Gamma$ that govern the system~\cite{NoiraySchu13,NoirayDenisov16}, as shown in figure~\ref{fig:extrap_3D}$c$.
Hereafter, we will denote the  KM coefficients \textit{estimated} from measurements with a hat~$\,\,\widehat{.}\,\,$, as opposed to the KM coefficients \textit{calculated} with the AFP equation (no hat).

\subsection{Adjoint-based optimisation}
\label{sec:exact}

\subsubsection{The adjoint Fokker-Planck equation}
\label{sec:AFPE}

This section presents an alternative method to compute  the KM coefficients $D^{(n)}(A)$ which does not suffer from  finite-time effects and does not call for extrapolation.

\begin{figure}[] 
\psfrag{A}[r][][1][-90]{$a$}
\psfrag{D1}[r][][1][-90]{$D^{(n)}_\tau(A) = \dfrac{1}{n!\tau}P^\dag(A,\tau)$}
\psfrag{t}[t][]{$t$ [s]}
\psfrag{tau}[t][]{$\tau$ [s]}
\psfrag{Pa(0)}[b][]{$P^\dag(a,0)$} 
\psfrag{Pa(end)}[b][]{$P^\dag(a,\tau)$}
\centerline{
   \hspace{-0.1cm}
   \begin{overpic}[height=8.5cm,tics=10]{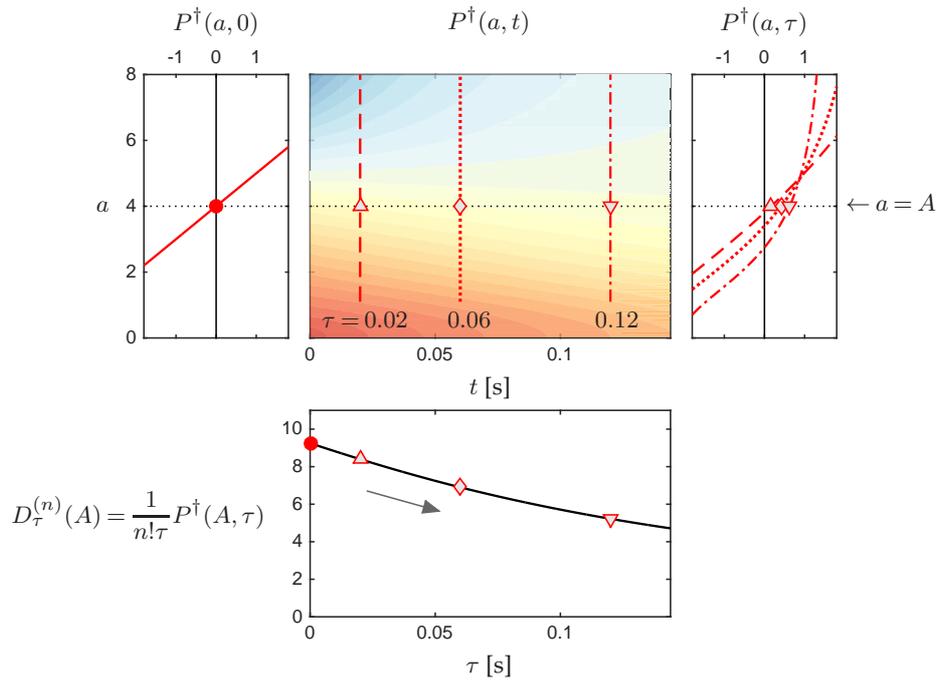}  
      \put(47,87.2){$P^\dag(a,t)$}      
      \put(101,62.4){\footnotesize $\leftarrow a=A$}
      \put(30,46.5){\footnotesize $\tau=0.02$}
      \put(47,46.5){\footnotesize      $0.06$}
      \put(67,46.5){\footnotesize      $0.12$}  
   \end{overpic}    
}
\caption{
Calculation of exact KM coefficients with the AFP equation~(\ref{eq:AFPE}):
starting from the initial condition $P^\dag(a,0) = (a-A)^n$, the solution $P^\dag(a,t)$ evaluated at later times $t=\tau$ and at the specific amplitude $a=A$ allows the computation of the exact KM coefficient $D_\tau^{(n)}(A) =  P^\dag(A,\tau) / (n! \tau)$.
The process is illustrated here with $n=1$, $A=4$, and $\tau=0.02$~s ($-\,\,\,-$), $0.06$~s ($\cdot\cdot\cdot$), $0.12$~s ($-\cdot-$). Note the absence of any finite-time effect in $\tau=0$. Parameters: $\nu=5$, $\kappa=2$, $\Gamma/4\omega_0^2=5$.
} 
\label{fig:AFP}
\end{figure}

Consider the adjoint Fokker-Planck (AFP) equation for $P^\dag(a,t)$:
\begin{equation}
\dfrac{\partial}{\partial t} P^\dag(a,t)
= 
D^{(1)} \dfrac{\partial}{\partial a}   P^\dag (a,t)  + 
D^{(2)}  \dfrac{\partial^2}{\partial a^2}   P^\dag(a,t).
\label{eq:AFPE}
\end{equation}%
Lade~\cite{LadePLA09} has shown that,  with a specific initial condition,
the solution of the AFP equation at $a=A$ and $t=\tau$ is related to the finite-time KM coefficient:
\begin{equation}
P^\dag(a,0) = (a-A)^n 
\quad \Rightarrow \quad P^\dag(A,\tau) = n! \tau D_\tau^{(n)}(A).
\label{eq:AFPE-init-final-cond}
\end{equation}
(See appendix~A for the derivation of (\ref{eq:AFPE})-(\ref{eq:AFPE-init-final-cond}).) Therefore, provided the KM coefficients $D^{(n)}(A)$  are known, the finite-time KM coefficients $D_\tau^{(n)}(A)$ can be calculated exactly for any $\tau$ by solving (\ref{eq:AFPE}) with the initial condition (\ref{eq:AFPE-init-final-cond}).
This is illustrated in figure~\ref{fig:AFP}: 
$D_\tau^{(1)}(4)$ is obtained by solving the AFP equation from the initial condition $P^\dag(a,0) = (a-4)^1$ and evaluating $P^\dag(4,\tau) / (1! \tau)$. This procedure is exact for any value of $\tau$ and is not affected by finite-time effects.

\begin{figure}[] 
\psfrag{A}[t][]{$A$}
\psfrag{DD} [r][][1][-90]{$D^{(1)}_\tau(A)$}
\psfrag{tau}[t][]{$\tau$~[s]}
\centerline{
   \begin{overpic}[height=6cm,tics=10]{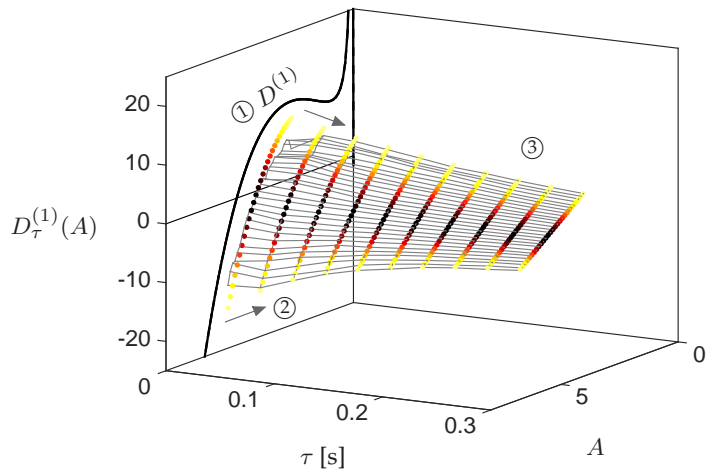}   
   \put( 22,55){\rotatebox{30}{${\mbox{\textcircled{\scriptsize 1}}}$  $D^{(1)}$} }  
   \put( 30,24){${\mbox{\textcircled{\scriptsize 2}}}$}  
   \put( 70,50){${\mbox{\textcircled{\scriptsize 3}}}$}  
   \end{overpic} 
}
\caption{
Adjoint-based system identification.
The exact KM coefficients $D^{(n)}(A)$  (${\mbox{\textcircled{\scriptsize 1}}}$) depend on the parameters $\{\nu,\kappa,\Gamma\}$ to be identified~(\ref{eq:D1D2}).
They are also involved in the exact calculation of finite-time KM coefficients $D_{\tau}^{(n)}(A)$  with the AFP equation~(\ref{eq:AFPE}) (${\mbox{\textcircled{\scriptsize 2}}}$).
Thus, system identification is made possible by adjusting $\{\nu,\kappa,\Gamma\}$ iteratively so as to minimize the overall error between estimated and calculated KM coefficients (${\mbox{\textcircled{\scriptsize 3}}}$).
Here $n=1$, $\nu=6$, $\kappa=2$, $\Gamma/4\omega_0^2=5$, $T=500$~s, $\Delta f=60$~Hz.
} 
\label{fig:AdjSI}
\end{figure}

In the context of system identification, the KM coefficients $D^{(n)}(A)$ are not known since they depend on the parameters  to be identified, $\{\nu,\kappa,\Gamma\}$.
However, combining the estimation of finite-time KM coefficients  $\widehat D_\tau^{(n)}(A)$ from measurements and the exact adjoint-based calculation of finite-time KM coefficients $D_\tau^{(n)}(A)$ 
yields a powerful system identification method \cite{Honisch11}:
given sets of  amplitudes and time shifts, adjust  $\{\nu,\kappa,\Gamma\}$ iteratively so as to  minimize the overall difference between $\widehat D_\tau^{(n)}(A)$ and  $D_\tau^{(n)}(A)$ (figure~\ref{fig:AdjSI}).
The strength of this method is twofold: extrapolation is not needed, and data at all amplitudes and time shifts are used simultaneously.
Details about the optimisation procedure are given in section~\ref{sec:SI}\ref{sec:exact}\ref{sec:num_meth}.

\subsubsection{Optimisation procedure and numerical method}
\label{sec:num_meth}

\begin{figure}[] 
\psfrag{11}[][]{Choose time shifts $\tau_i$ }
\psfrag{12}[][]{and amplitudes $A_j$}
\psfrag{21}[][]{${\mbox{\textcircled{\scriptsize 0}}}$
Estimate finite-time KM coefficients }
\psfrag{22}[][]{$\widehat D^{(n)}_{\tau_i}(A_j)$ from time signal $A(t)$}
\psfrag{31}[][]{Set initial values $\{\nu_0,\kappa_0,\Gamma_0\}$}
\psfrag{32}[][]{for the parameters to be identified}
\psfrag{41}[][]{${\mbox{\textcircled{\scriptsize 1}}}$ 
Using the current $\{\nu,\kappa,\Gamma\}$ (current $D^{(n)}(A)$),
}
\psfrag{42}[][]{${\mbox{\textcircled{\scriptsize 2}}}$ compute finite-time KM coefficients}
\psfrag{43}[][]{$D^{(n)}_{\tau_i}(A_j)$ with the AFP equation}
\psfrag{51}[][]{Evaluate the overall error $\mathcal{E}$}
\psfrag{52}[][]{between $\widehat D^{(n)}_{\tau_i}(A_j)$ and $D^{(n)}_{\tau_i}(A_j)$}
\psfrag{61}[][]{\footnotesize $ \quad \qquad \qquad \qquad$ If $\mathcal{E} \geq$ tolerance}
\psfrag{62}[][]{\footnotesize $ \quad \qquad \qquad \qquad$ If $\mathcal{E} \leq$ tolerance}
\psfrag{71}[][]{Update $\{\nu,\kappa,\Gamma\}$}
\psfrag{72}[][]{SI converged}
\centerline{
   \begin{overpic}[height=9cm,tics=10]{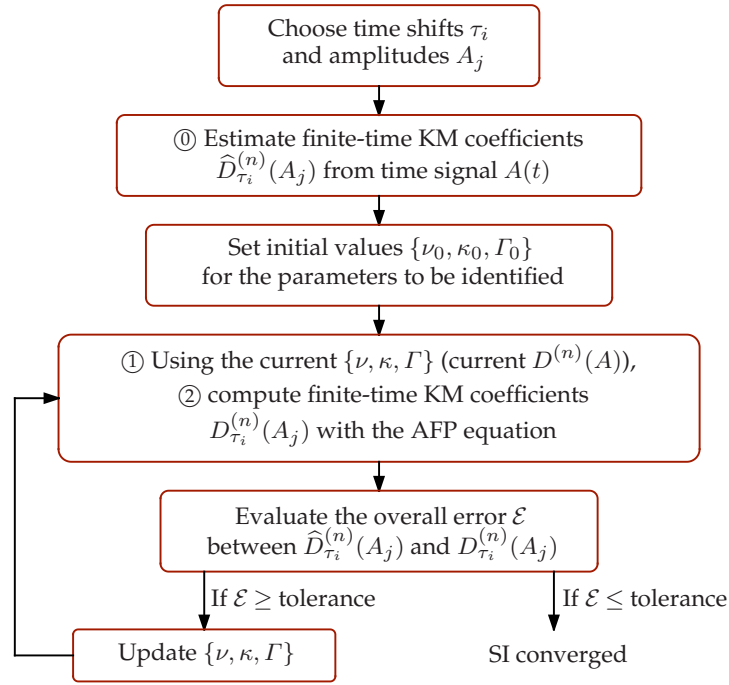}       
   \end{overpic}    
}
\caption{
Adjoint-based system identification.
Circled numbers refer to figure~\ref{fig:AdjSI}.
}
\label{fig:blockdiagram}
\end{figure}

Optimisation is performed as detailed below (see also figure~\ref{fig:blockdiagram}). Given a time signal $A(t)$ in the stationary regime:
\begin{itemize}
\item
Choose a set of  $N_\tau$  
time shifts $\tau_i$,  and  $N_A$  
amplitudes $A_j$, ($i=1 \ldots N_\tau$, $j=1 \ldots N_A$);
\item 
Estimate the finite-time KM coefficients $\widehat D^{(n)}_{\tau_i}(A_j)$ from the time signal, as described in section~\ref{sec:finiteKM}.
(This step is the same  in the extrapolation-based SI method described earlier and in the present adjoint-based SI method.)
Although not indispensable, subsequent kernel-based regression yields smoother data~\cite{Honisch11};
\item
Choose a set of initial values  for the parameters of interest, here $\{\nu,\kappa,\Gamma\}=\{\nu_0,\kappa_0,\Gamma_0\}$;
\item
Compute the finite-time KM coefficients $D^{(n)}_{\tau_i}(A_j)$ by 
solving the AFP equation (\ref{eq:AFPE})
 $2 N_A$ times with a different initial condition  (\ref{eq:AFPE-init-final-cond}) for each amplitude $A_j$, and $n=1,2$. 
Here the exact KM coefficients $D^{(n)}(A)$ of the AFP operator are evaluated according to (\ref{eq:D1D2}) with the current parameter values $\{\nu,\kappa,\Gamma\}$;
\item
Compute the weighted error between  estimated and calculated finite-time KM coefficients
\begin{equation}
  \mathcal{E}( \nu,\kappa,\Gamma )
= 
  \dfrac{1}{2 N_\tau N_A}
  \sum_{n=1}^{2} 
  \sum_{i=1}^{N_\tau} 
  \sum_{j=1}^{N_A} 
W_{ij}^{(n)}  \left(  \widehat D_{\tau_i}^{(n)}(A_j) -  D_{\tau_i}^{(n)}(A_j; \nu,\kappa,\Gamma ) \right)^2;
\label{eq:error}
\end{equation}
\item
Modify the parameters $\{\nu,\kappa,\Gamma\}$ so as to reduce the error; solve again the AFP equation;  iterate until convergence is reached, thus effectively solving the problem 
\begin{equation}
\min_{\{\nu,\kappa,\Gamma\}} \mathcal{E}.
\end{equation}
\end{itemize}

In our implementation, the time shifts $\tau_i$ are distributed uniformly in the interval $[\tau_1,\tau_2]$, chosen so that the estimated KM coefficients are both relevant ($\tau_1$ is large enough to avoid finite-time effects) and useful ($\tau_2$ is  small enough for $A(t)$ and $A(t+\tau_i)$ to have non-zero correlation). 
Specifically, the lower bound $\tau_1$ is chosen as 
$\max \{ f_s^{-1} , \Delta f^{-1}, (\omega_0/2\pi)^{-1} \}$, 
where $f_s$ is the measurement sampling frequency, and $\Delta f$ is the filtering bandwidth (that introduces a finite correlation of the envelope).
The upper bound $\tau_2$ is set to $2 \tau_A$, where $\tau_A$ is such that the autocorrelation function of the envelope drops significantly,  $k_{AA}(\tau_A)=k_{AA}(0)/4$.

The AFP equation is solved on $(a,t) \in
[0,a_{\infty}] \times [0,\tau_2]$ with a second-order finite-difference discretisation in space and a second-order Crank-Nicolson discretisation in time.
The boundary $a_{\infty}$ is set to 1.5 times the largest amplitude measured in the signal, $\max_t(A(t))$. The numerical method, implemented in \textit{Matlab}, has been validated with available analytical solutions~\cite{Honisch11}. The $2N_A$ simulations are independent and can therefore be run very efficiently in parallel.

The weights $W_{ij}^{(n)}$ introduced in the error function $\mathcal{E}$ serve a twofold purpose:
(i)~account for the higher statistical accuracy for amplitudes of higher probability, and
(ii)~normalize the first-order and second-order KM coefficients to ensure that their relative contributions are of the same order of magnitude. To this aim, we choose weights 
$W_{ij}^{(n)} = (n! \tau_i)^2 P(A_j) / V^{(n)}$
that include (i)~the PDF $P(A_j)$ itself, and (ii)~the variance of the PDF-weighted KM coefficients $V^{(n)} = \mbox{Var}_{i,j} \{ n! \tau_i P(A_j)  \widehat D_{\tau_i}^{(n)}(A_j) \}$. 

At each iteration, parameters $\{\nu,\kappa,\Gamma\}$ are updated using a simplex search algorithm~\cite{Lagarias98}.
Convergence is reached when all absolute and relative  variations in $\{\nu,\kappa,\Gamma/4\omega^2\}$, as well as in $\mathcal{E}$, are smaller than $10^{-2}$.
Note that the optimisation procedure can be repeated from different initial values $\{\nu_0,\kappa_0,\Gamma_0\}$ in order to assess  whether the obtained local minimum is likely to be global.

\subsubsection{Example}
\label{sec:valid_num}

\begin{figure}[] 
\psfrag{iter}[t][]{iterations}
\psfrag{J}[r][][1][-90]{$\mathcal{E}$}
\psfrag{nu}[r][][1][-90]{$\nu$}
\psfrag{nv}[r][][1][-90]{}
\psfrag{kappa}[t][]{$\kappa$}
\psfrag{Gamma}[t][]{$\Gamma/4\omega^2$}
\psfrag{.}[t][]{}
\centerline{
   \hspace{0.05cm}
   \begin{overpic}[height=4.2cm,tics=10]{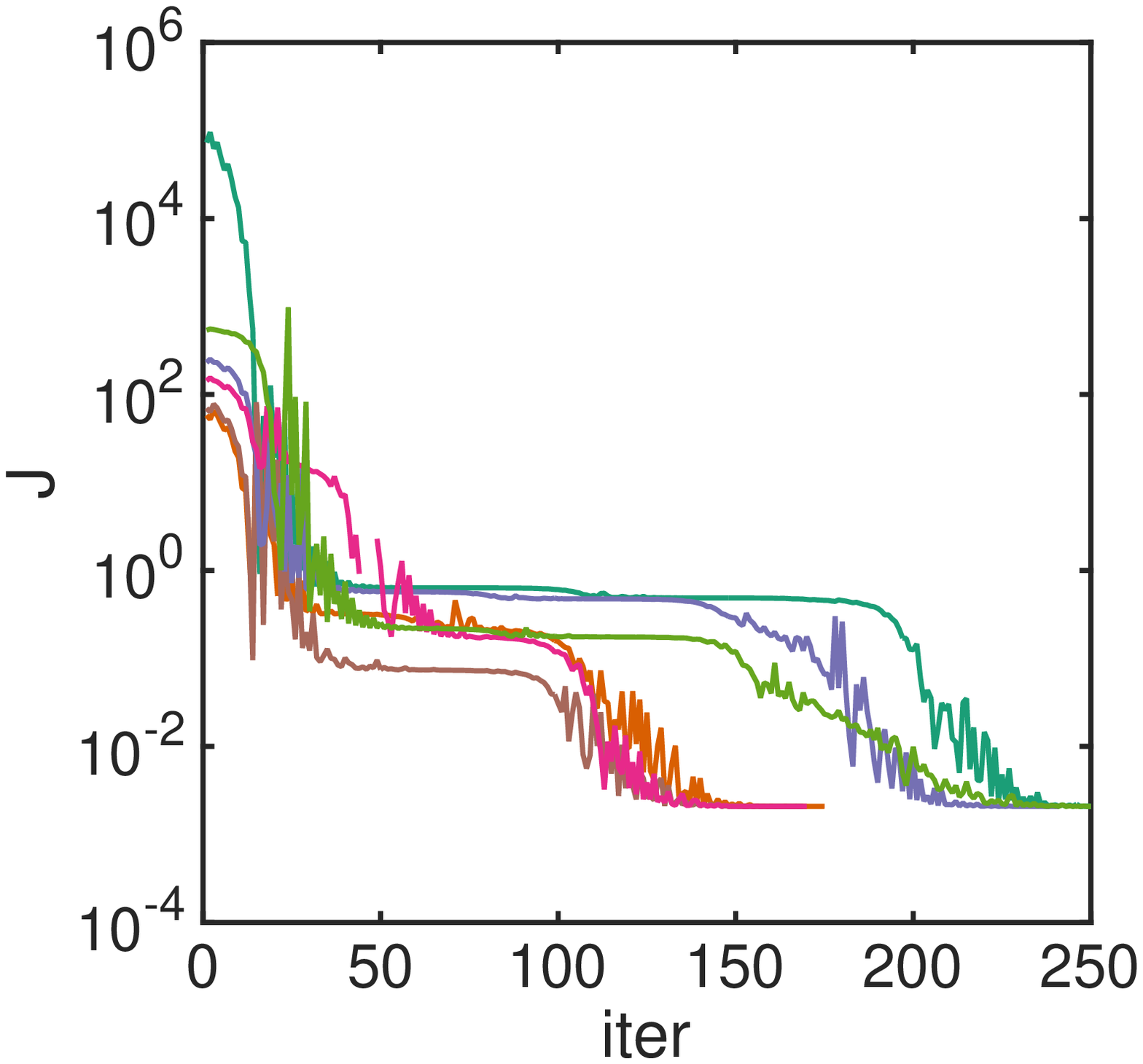}  
      \put(-5,87){$(a)$}      
   \end{overpic}  
   \hspace{0.4cm}  
   \begin{overpic}[height=4.2cm,tics=10]{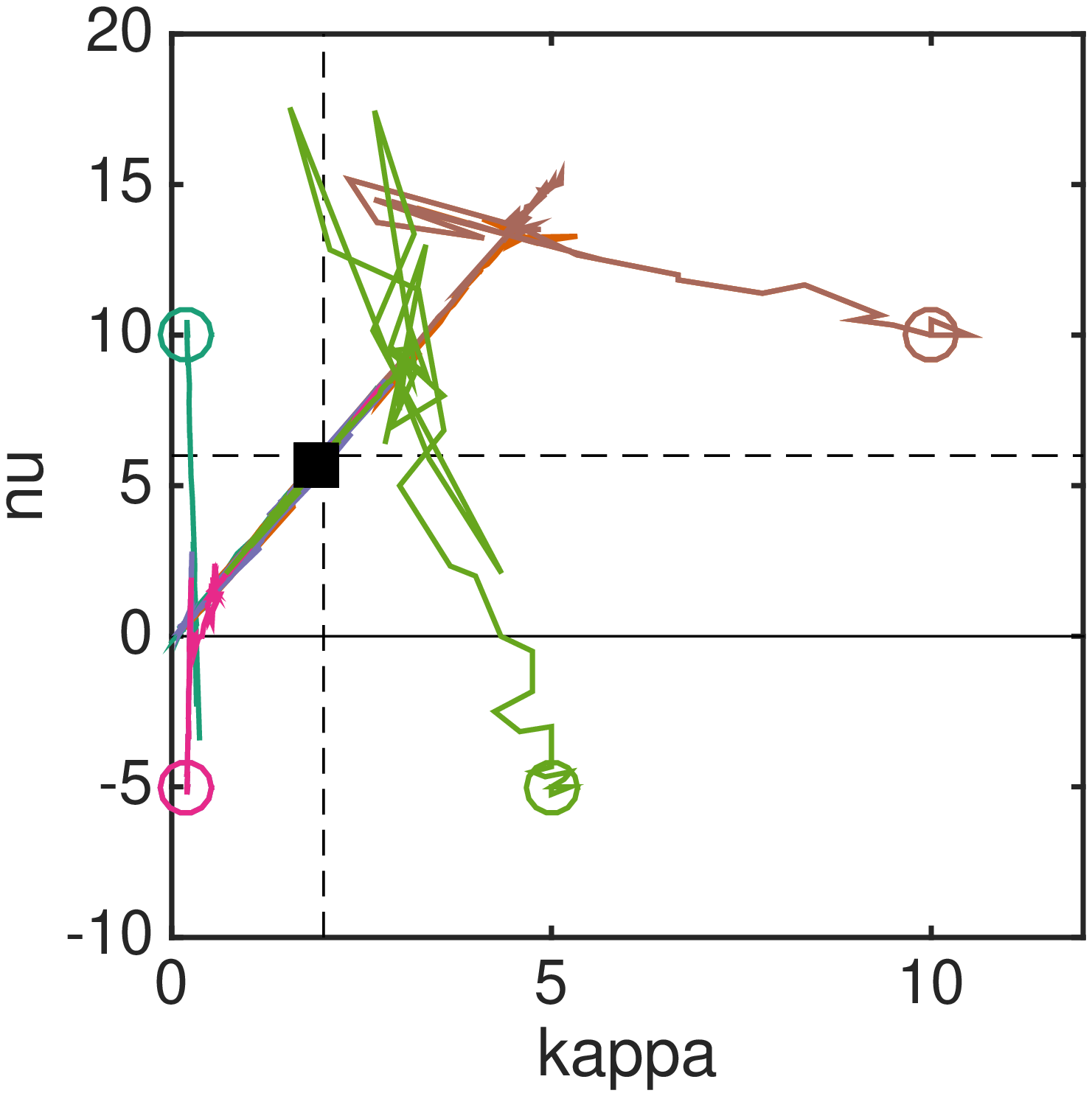}  
      \put(-5,92){$(b)$} 
   \end{overpic}    
   \hspace{0.17cm}  
   \begin{overpic}[height=4.2cm,tics=10]{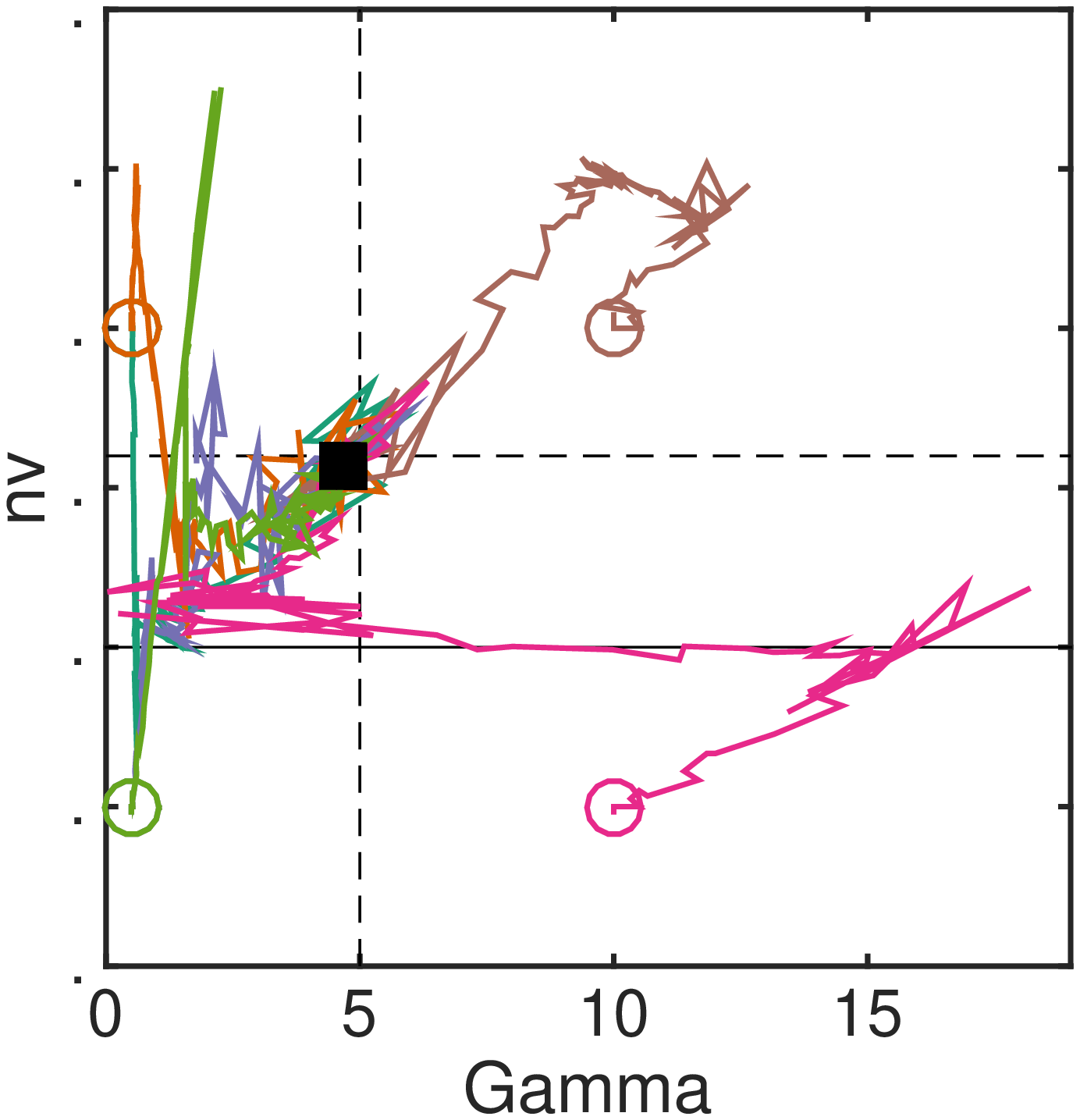}  
   \end{overpic}    
}
\caption{
Convergence history of the adjoint-based optimisation starting from 6 different initial values $\{\nu_0,\kappa_0,\Gamma_0\}$.
$(a)$~Error (\ref{eq:error}).
$(b)$~Parameters $\{ \nu,\kappa,\Gamma/4\omega^2 \}$. Circles show initial values, squares show final values. Dashed lines show the exact values $\nu=6$, $\kappa=2$, $\Gamma/4\omega^2=5$. 
Other parameters: $\omega_0/2\pi=150$~Hz, $T=500$~s, $\Delta f=60$~Hz.
} 
\label{fig:histparams}
\end{figure}

We apply the adjoint-based system identification method to synthetic signals. The VdP oscillator~(\ref{eq:eta3}) is simulated with \textit{Simulink}, using  $\nu=6$, $\kappa=2$, $\Gamma/4\omega^2=5$, $\omega_0/2\pi=150$~Hz, $T=500$~s. The pressure signal $\eta(t)$ in the stationary regime is band-pass filtered around $\omega_0/2\pi$ with bandwidth $\Delta f=60$~Hz. The envelope $A(t)$ is extracted with the Hilbert transform. Finite-time KM coefficients are estimated and calculated with $N_\tau=10$ and $N_A=49$. (The influence of several of these parameters is reported in appendix~B.)

\begin{figure}[] 
\psfrag{A}[t][]{$A$}
\psfrag{D} [r][][1][-90]{$D^{(1)}_\tau(A)$}
\psfrag{D1}[r][][1][-90]{$D^{(1)}_\tau(A)$}
\psfrag{DD}[r][][1][-90]{}
\psfrag{tau}[t][]{$\tau$~[s]}
\centerline{
   \begin{overpic}[height=6cm,tics=10]{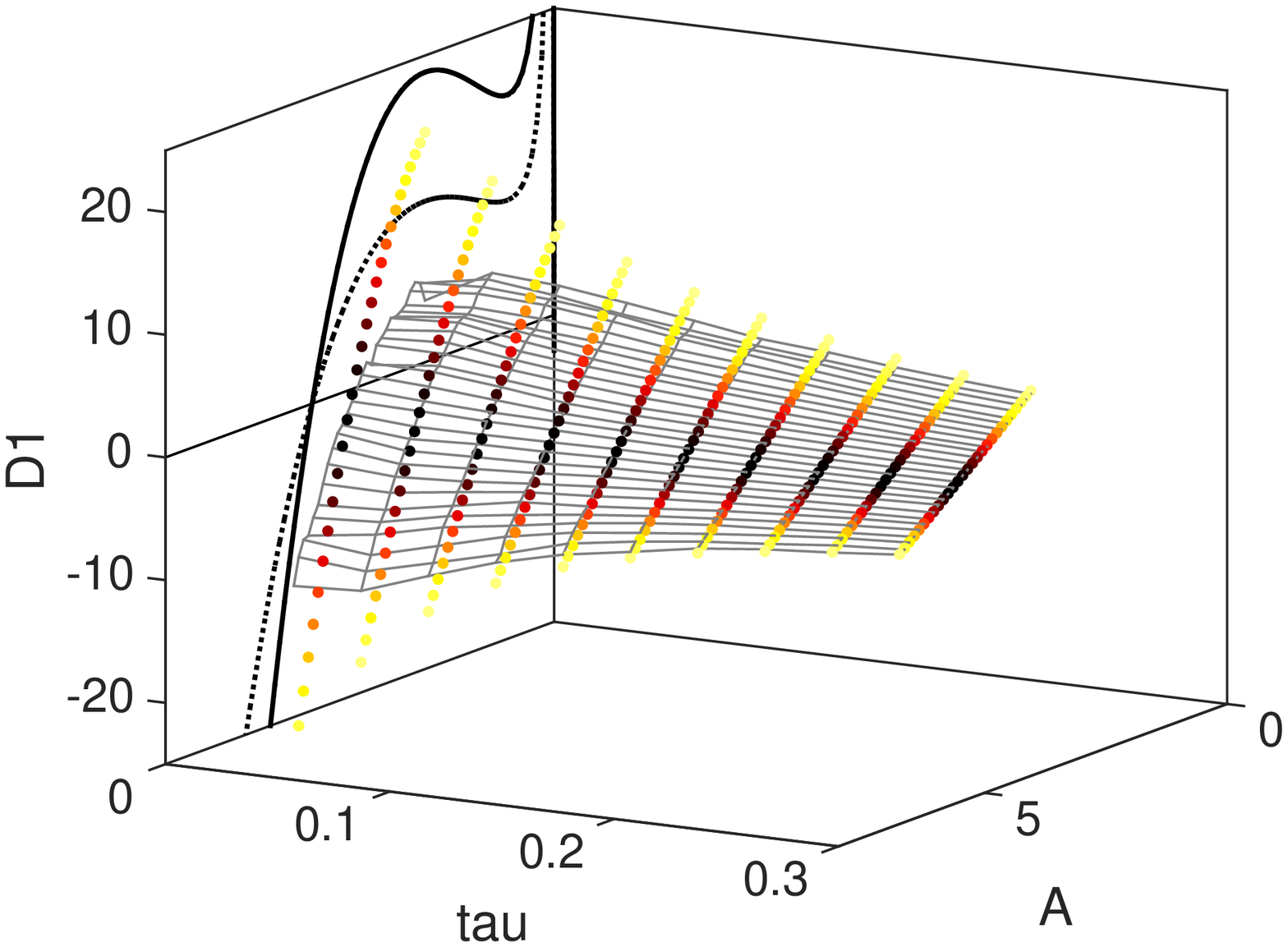}       
   \end{overpic}  
}
\vspace{0.2cm}
\centerline{   
   \begin{overpic}[height=6cm,tics=10]{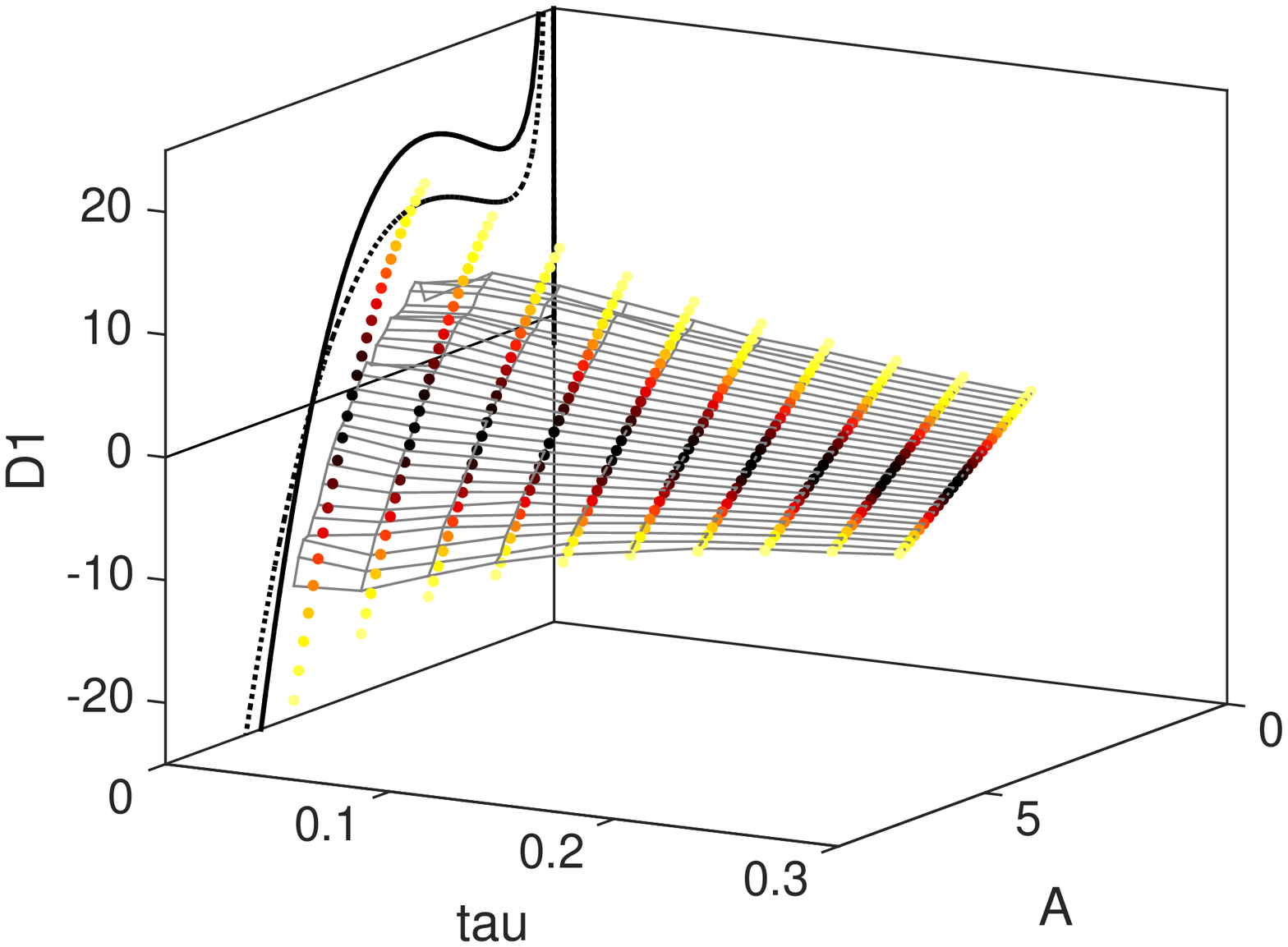}       
   \end{overpic}  
}
\vspace{0.2cm}
\centerline{  
   \begin{overpic}[height=6cm,tics=10]{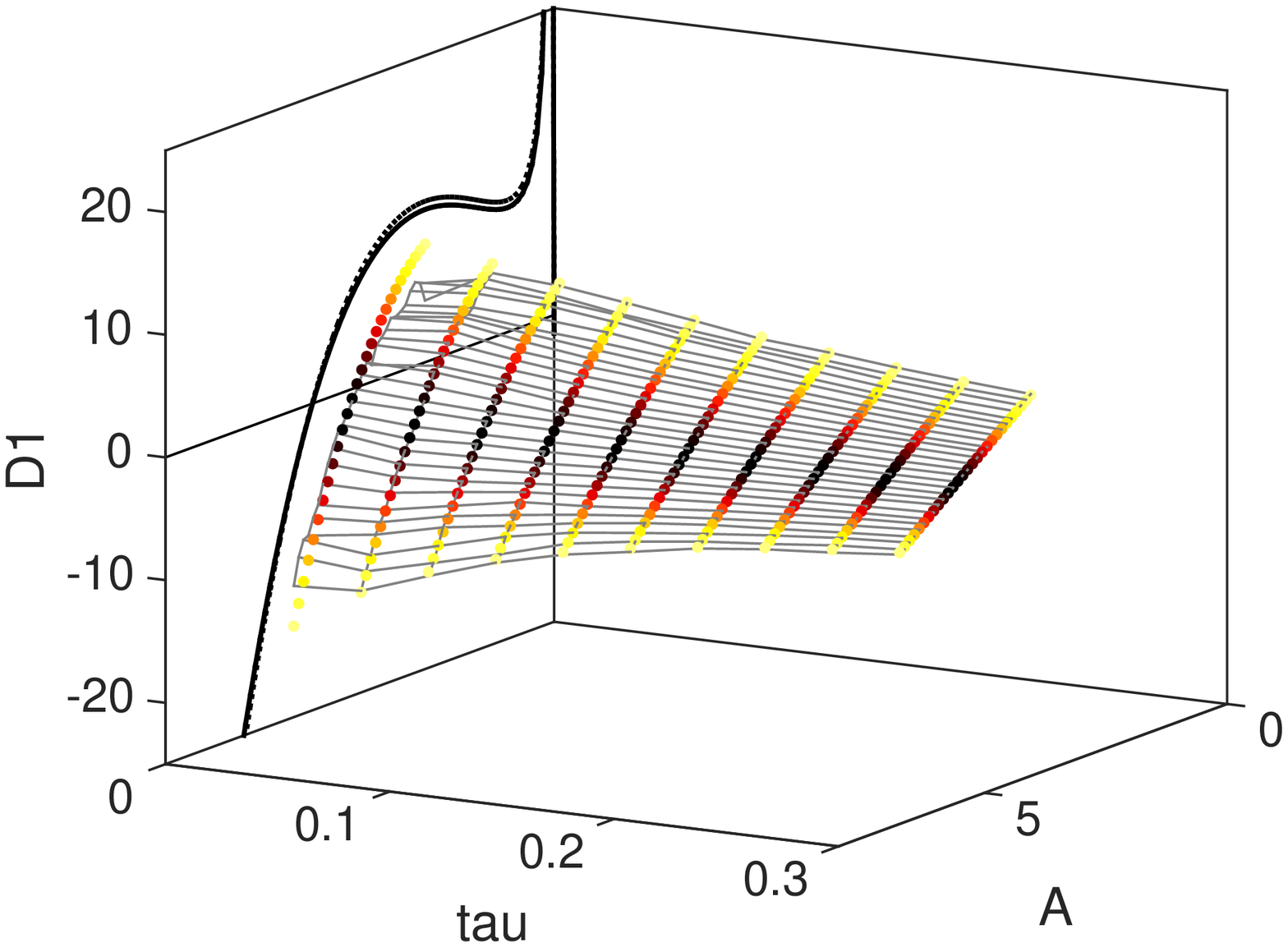}  
   \end{overpic}  
}
\caption{
Convergence history of the adjoint-based optimisation: finite-time KM coefficients $\widehat D_\tau^{(1)}$ estimated once from the time signal $A(t)$ (surface), and $D_\tau^{(1)}$ calculated with the AFP equation using different parameter values $\{\nu,\kappa,\Gamma\}$ at each iteration (dots).
$(a,b)$~Intermediate iterations, $(c)$~final iteration. At $\tau=0$, the dashed and solid lines show the exact analytical $D^{(1)}$ and the current tentative $D^{(1)}$, respectively.
Parameters: $\omega_0/2\pi=150$~Hz, $T=500$~s, $\Delta f=60$~Hz.
} 
\label{fig:histD}
\end{figure}

Figure~\ref{fig:histparams} shows the convergence history obtained for the same signal when starting with 6 different initial values $\{\nu_0,\kappa_0,\Gamma_0\}$. In all cases the error decreases by several orders of magnitude (panel $a$), and after different  paths in the parameter space the same set of parameters is identified close to the exact values (panel $b$). This shows that the method is both robust and accurate.
Estimated and calculated finite-time KM coefficients  are shown at different iterations in figure~\ref{fig:histD}. 
One can clearly see how adjusting the parameters $\{\nu_0,\kappa_0,\Gamma_0\}$  gradually reduces the error for all time shifts $\tau$ and amplitudes $A_j$.

\section{Conclusion}

In this paper we have proposed an output-only system identification method for extracting the parameters governing stochastic harmonic oscillators. Using the adjoint Fokker-Planck equation yields a method unaffected by finite-time effects, unlike the direct evaluation of the Kramers-Moyal coefficients. Accuracy and robustness are provided by performing a global optimisation over a whole range of amplitudes and time shifts.
Establishing an explicit link between the Fokker-Planck equation and the second-order oscillator's stochastic differential equation (rather than the first-order stochastic amplitude equation) allows for the identification of the physical parameters of the system such as linear growth rate and nonlinear damping.

One could think of choosing the oscillation variable $\eta(t)$ as an alternative to the  envelope $A(t)$, which would require handling a two-dimensional Fokker-Planck equation for $(\eta,\dot\eta)$.
Note also that we have focused on an individual oscillator by band-pass filtering the time signal around the frequency of interest; 
future efforts should investigate the possibility to apply the present adjoint-based system identification method simultaneously to several oscillators with nearby frequencies, which would also involve a multi-dimensional Fokker-Planck equation.
Another topic of interest is the presence of stiffness nonlinearities, such as in the Duffing oscillator and Duffing-Van der Pol oscillator;
in this case, the amplitude and phase dynamics are fully coupled and one should  consider a suitable two-dimensional Fokker-Planck equation. These few examples show that although adjoint-based system identification would become more involved, multi-dimensional Fokker-Planck equations would allow for valuable progress.

\vskip6pt

\enlargethispage{20pt}




\aucontribute{
N.N. conceived of the study. 
E.B. implemented the method, carried out numerical simulations and analysed the data. 
Both authors drafted the manuscript.
}

\competing{
We declare we have no competing interests.
}

\funding{
The authors acknowledge support from Repower and the ETH Zurich Foundation.
}

\ack{
The authors thank A. H\'ebert for his work during the initial stage of the study.
}


\section*{Appendix A. Derivation of the adjoint Fokker-Planck equation}
\label{AppA}

We include for completeness the derivation of the AFP equation (\ref{eq:AFPE}) and of relation (\ref{eq:AFPE-init-final-cond}) for the exact calculation of finite-time KM coefficients, following closely Honisch and Friedrich~\cite{Honisch11} and Lade~\cite{LadePLA09}.
Define the Fokker-Planck operator $\mathcal{L}(A) =  - \partial_A D^{(1)}(A)  +  \partial_{AA} D^{(2)}(A)$ and consider again the FP equation (\ref{eq:FPE}) for  $P(A,t)$,
\begin{align}
\frac{\partial }{\partial t} P(A,t)
= \mathcal{L}(A) \, P(A,t),
\label{eq:FPE_new1}
\end{align}
whose solution reads formally
\begin{align}
 P(A,t) = e^{\mathcal{L}(A) t} P(A,0),
\label{eq:FPE_new2}
\end{align}
with the classical definition of  the exponential operator $e^{\mathcal{L}(A)t} = \sum_{k=0}^\infty \frac{1}{k!} \left( \mathcal{L}(A) t \right)^k$.
Recall that the conditional PDF is also solution of the  FP equation,
\begin{align}
\frac{\partial }{\partial \tau} P(a,t+\tau|A,t)
= \mathcal{L}(a) \, P(a,t+\tau|A,t),
\label{eq:FPE_new3}
\end{align}
and can therefore be expressed as
\begin{align}
P(a,t+\tau|A,t) = e^{\mathcal{L}(a) \tau} P(a,t+0|A,t)
= e^{\mathcal{L}(a) \tau} \delta(a-A).
\label{eq:FPE_new4}
\end{align}
Inserting in the definition (\ref{eq:D}) of the finite-time KM coefficient yields
\begin{eqnarray}
n!\tau D^{(n)}_\tau(A) &=& \int_{0}^{\infty} (a-A)^n P(a,t+\tau|A,t) \,\mathrm{d}a
\nonumber
\\
&=&
\int_{0}^{\infty} (a-A)^n   \left[ e^{\mathcal{L}(a) \tau} \delta(a-A) \right] \,\mathrm{d}a
\nonumber
\\
&=&
\int_{0}^{\infty} \left[   e^{\mathcal{L^\dag}(a)  \tau} (a-A)^n \right]  \delta(a-A) \,\mathrm{d}a
\nonumber
\\
&=&
\left.   e^{\mathcal{L^\dag}(a)  \tau} (a-A)^n \right|_{a=A},
\label{eq:detailAFP}
\end{eqnarray}
where the AFP operator $ \mathcal{L^\dag}(a) = D^{(1)}(a) \partial_a   + D^{(2)}(a)  \partial_{aa}$  associated with the scalar product
$\left(u \, | \, v \right) = \int_0^\infty u(a) v(a) \,
\mathrm{d}a$ is obtained with successive integrations by parts,
\begin{equation}
\int_{0}^{\infty} u(a)   \left[ \mathcal{L}(a) v(a) \right] \,\mathrm{d}a = \int_{0}^{\infty}  \left[ \mathcal{L}^\dag(a) u(a) \right]  v(a) \,\mathrm{d}a
\end{equation}
for any functions $u(a)$, $v(a)$ decaying to 0 in
$a=0$ and $a=\infty$ 
such that boundary terms vanish. (The singular case $A=0$ needs not be considered since $P(0,t)=0 \,\, \forall t$ and the FP equation is not useful for this particular value.)
With the same interpretation as in (\ref{eq:FPE_new1})-(\ref{eq:FPE_new4}),  relation (\ref{eq:detailAFP}) shows that $n!\tau D^{(n)}_\tau(A)$ is solution of the equation 
\begin{equation}
\dfrac{\partial}{\partial t} P^\dag(a,t)
= 
 \mathcal{L^\dag}(a) P^\dag(a,t)
\label{eq:AFPE_new1}
\end{equation}
solved with the initial condition
\begin{equation}
P^\dag(a,0) = (a-A)^n
\label{eq:AFPE_new2}
\end{equation}
and evaluated at $t=\tau$, $a=A$.
We thus recover (\ref{eq:AFPE}) and (\ref{eq:AFPE-init-final-cond}).

\section*{Appendix B. Robustness and accuracy of the adjoint-based system identification}
\label{AppB}

\begin{figure}[] 
\psfrag{nu}[t][]{$\nu$}
\psfrag{BW}[t][]{$\Delta f$}
\psfrag{T}[t][]{$T$}
\psfrag{nuth}[t][]{Exact $\nu$}
\psfrag{nuopt,kopt,Gopt} [][]{Identified parameters}
\psfrag{A}[r][][1][-90]{$A$}
\centerline{
   \begin{overpic}[height=5.3cm,tics=10]{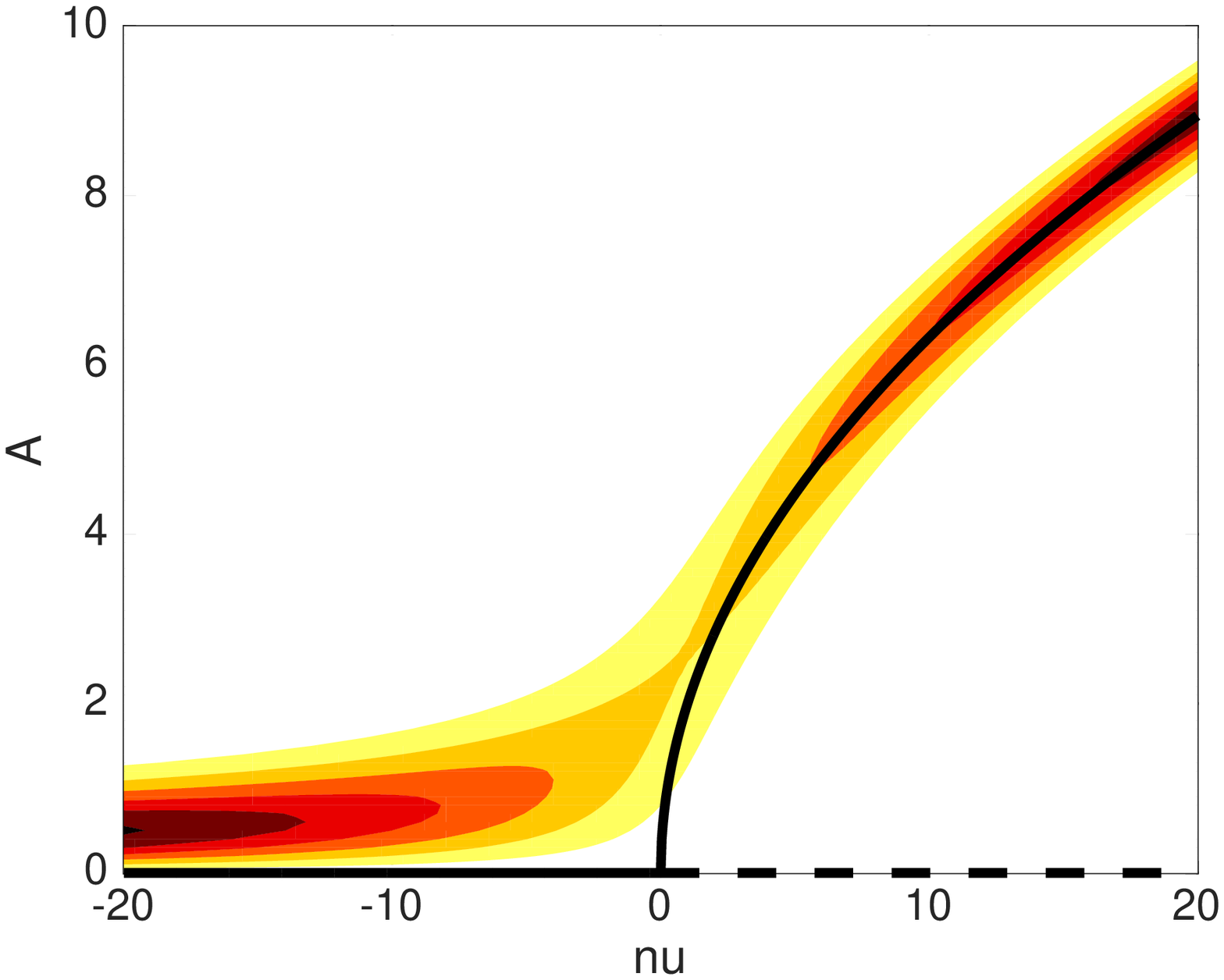}   
      \put(-3,74){$(a)$}   
   \end{overpic}
   \hspace{0.3cm}  
   \begin{overpic}[height=5.3cm,tics=10]{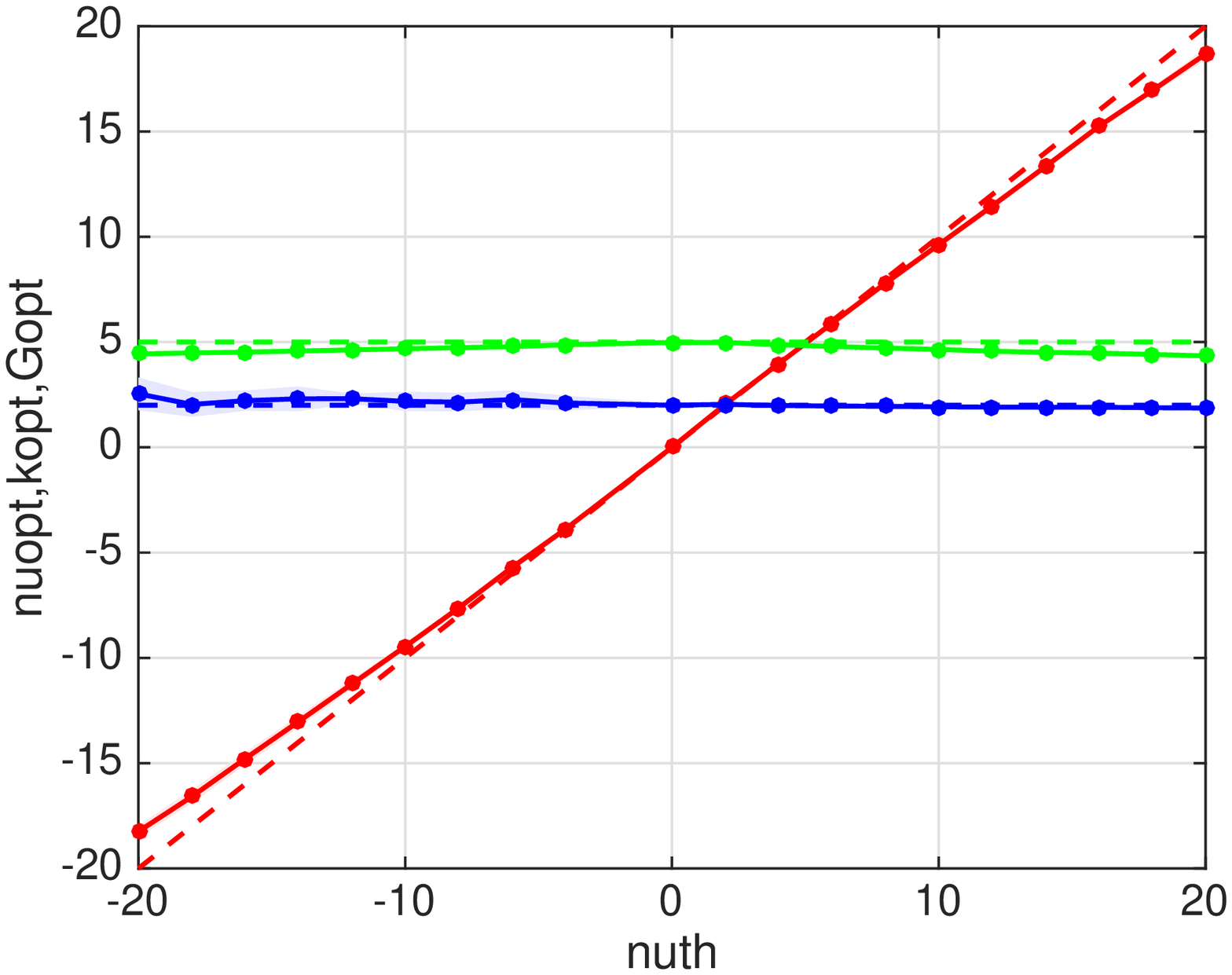}   
      \put(-1.5,74){$(b)$}  
      \put(15,55){ \textcolor[rgb]{0,0.7,0}{$\Gamma/4\omega_0^2$} } 
      \put(15,42){ \textcolor{blue}{$\kappa$} } 
      \put(15,20){ \textcolor{red}{$\nu$} } 
   \end{overpic}  
}
\vspace{0.6cm}
\centerline{
   \begin{overpic}[height=5.3cm,tics=10]{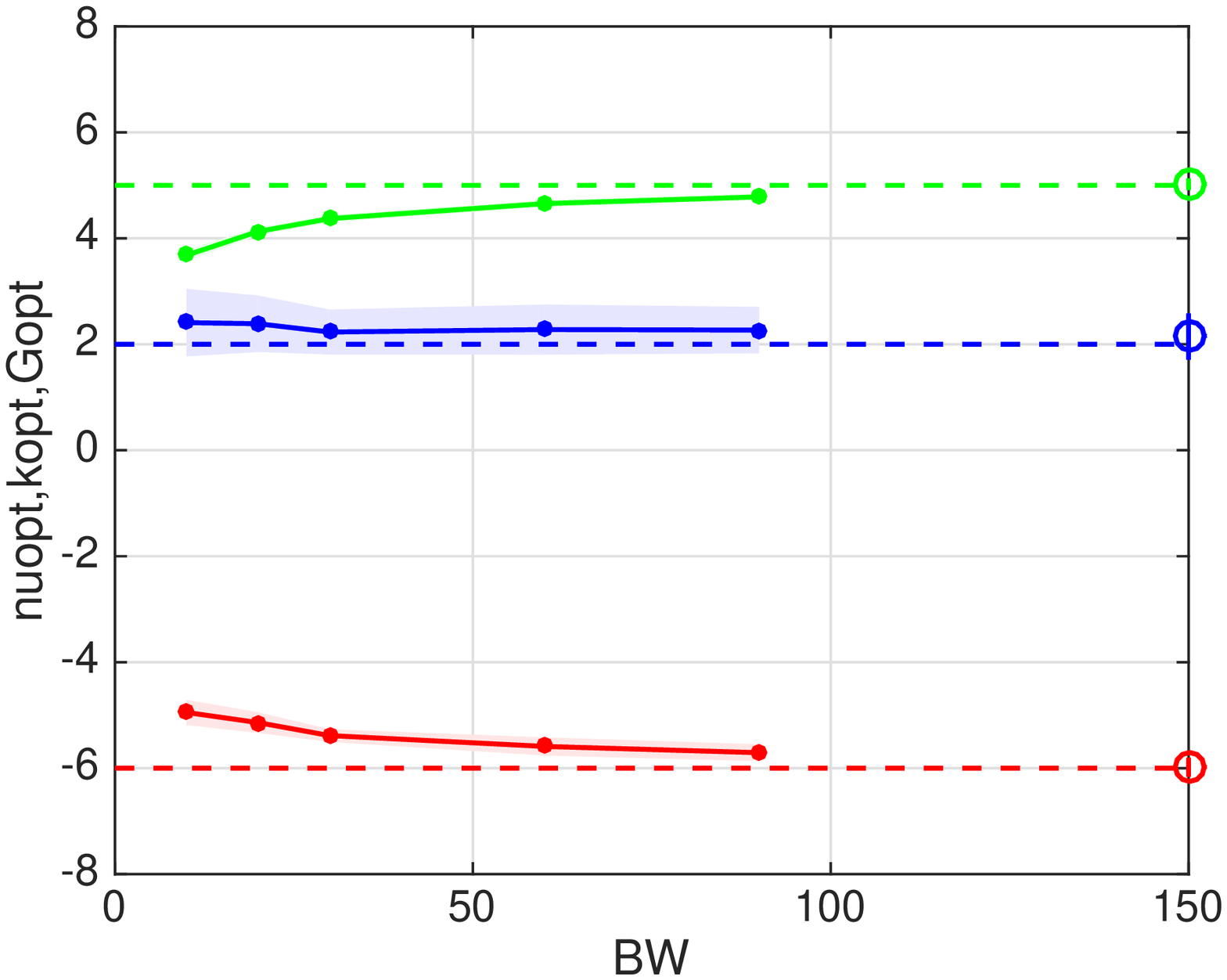}  
      \put(-1.5,74){$(c)$}    
   \end{overpic}  
   \hspace{0.4cm}
   \begin{overpic}[height=5.3cm,tics=10]{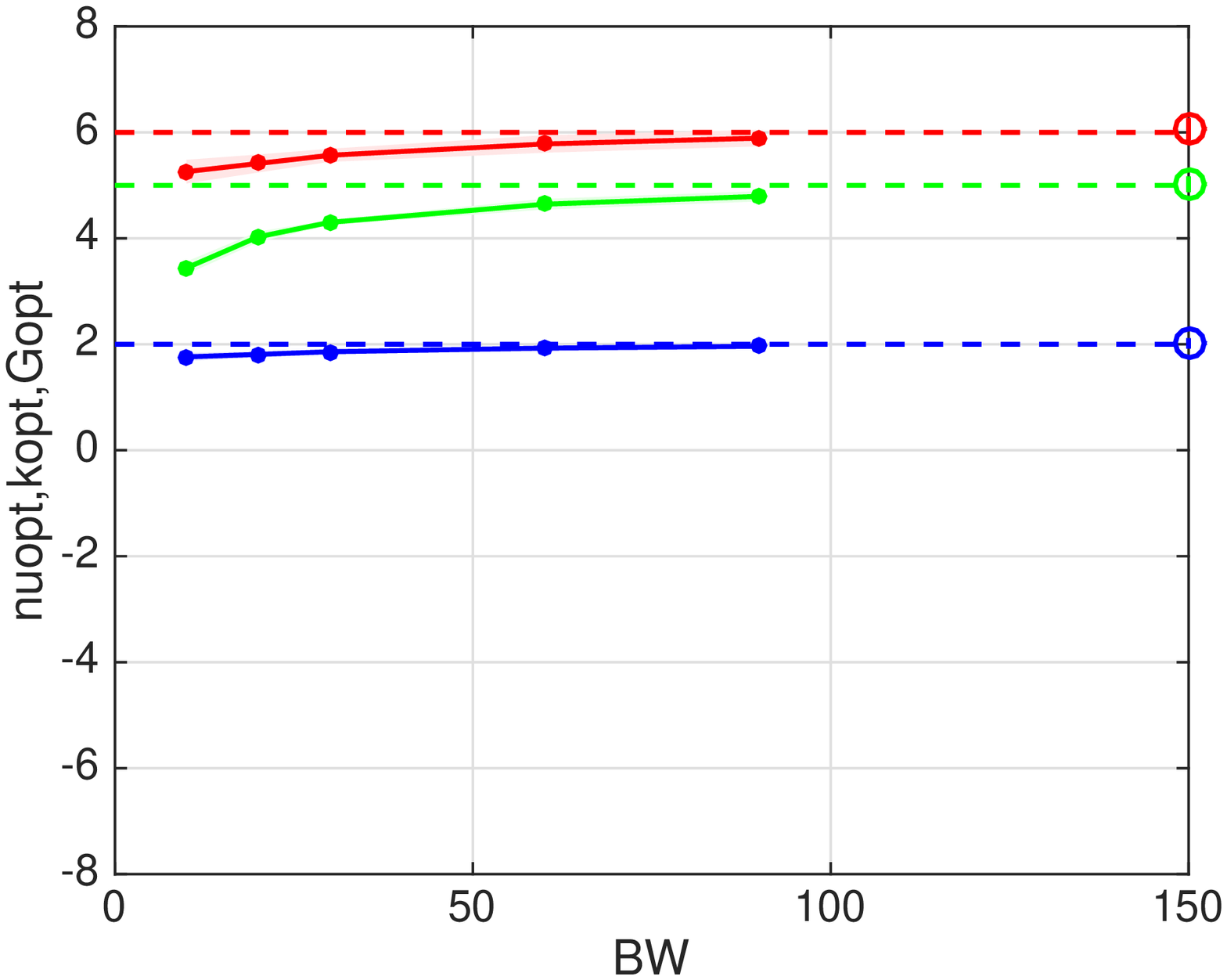}      
   \end{overpic} 
}
\vspace{0.6cm}
\centerline{
   \begin{overpic}[height=5.3cm,tics=10]{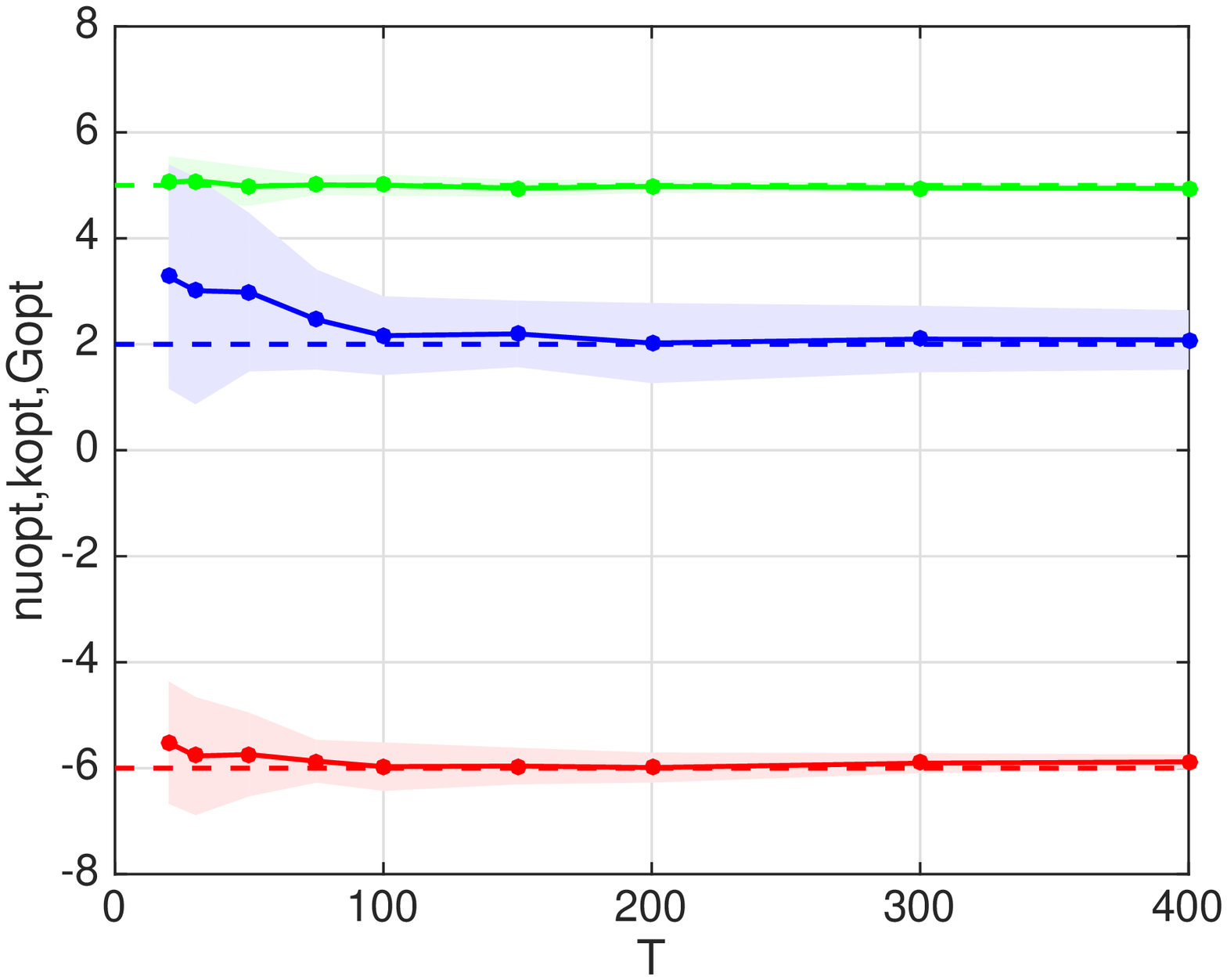}
      \put(-1.5,74){$(d)$} 
   \end{overpic}  
   \hspace{0.4cm}
   \begin{overpic}[height=5.3cm,tics=10]{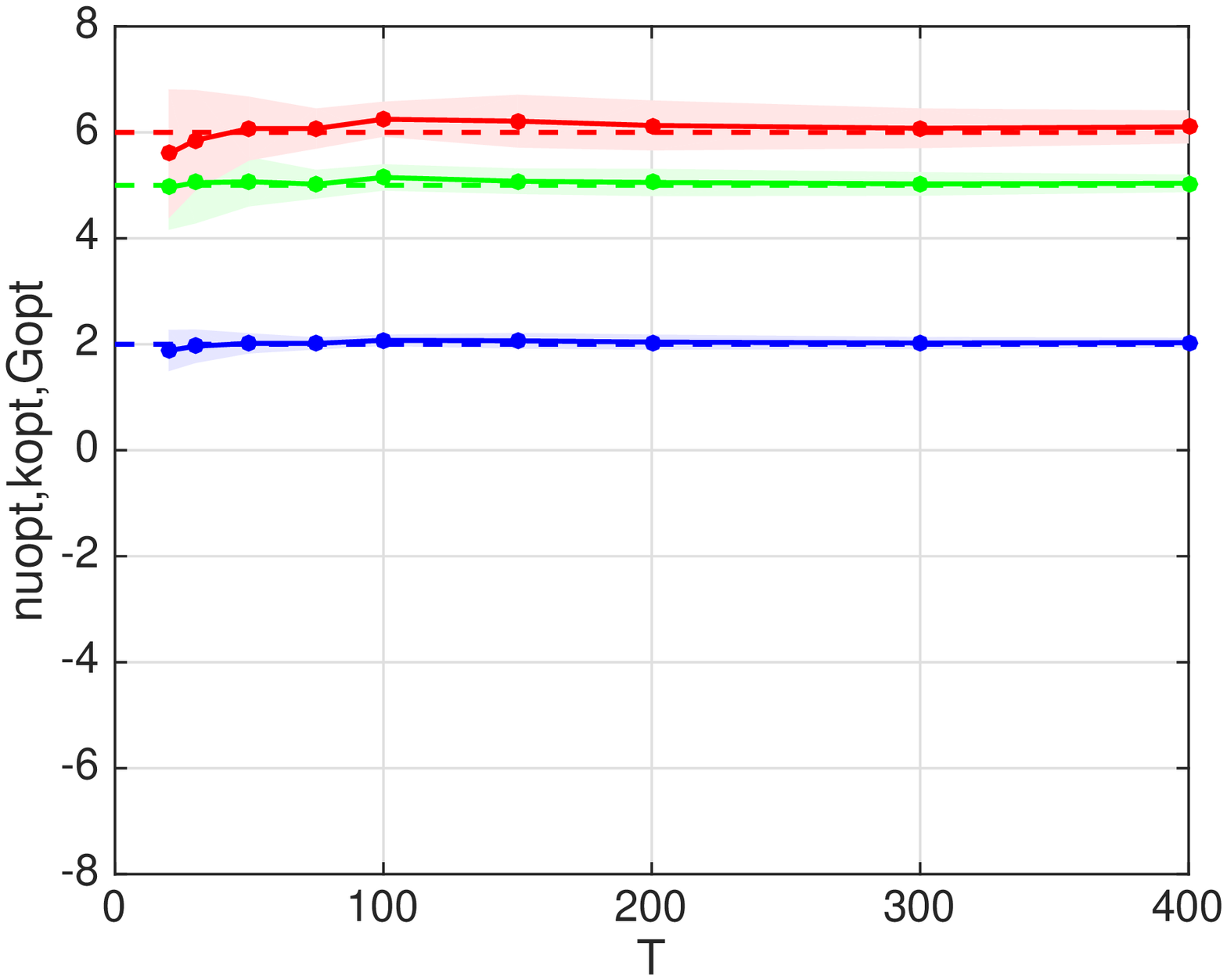}      
   \end{overpic} 
}
\caption{
$(a)$~Bifurcation diagram: deterministic amplitude (solid line, $A_{det}=0$ when $\nu<0$, $A_{det}=\sqrt{8\nu/\kappa}$ when $\nu>0$) and PDF $P(A;\nu)$ (contours).
$(b)-(d)$~Adjoint-based system identification. Dots: mean value (from 10 simulations in each configuration); shaded areas: standard deviation; dashed lines: exact values.
$(b)$~Identified parameters when varying the exact growth rate, $-20 \leq \nu \leq 20$ ($T=1000$~s, $\Delta f=90$~Hz);
$(c)$~Effect of the filtering bandwidth $\Delta f$ ($\nu=-6$ and $6$, $T=1000$~s). Circles on the rightmost side: no filtering;
$(d)$~Effect of the signal duration $T$ ($\nu=-6$ and $6$, no filtering). In all cases, $\kappa=2$, $\Gamma/4\omega^2=5$, $\omega_0/2\pi=150$~Hz.
} 
\label{fig:sweeps}
\end{figure}

Figure~\ref{fig:sweeps} presents the results of the adjoint-based system identification obtained  for various sets of parameters.
The exact growth rate $\nu$ is varied from -20 to 20, and $\kappa=2$, $\Gamma/4\omega^2=5$, $\omega_0/2\pi=150$~Hz.
The deterministic and stochastic bifurcation diagrams for these parameters are shown in panel $(a)$.
The mean value and standard deviation (calculated from 10 independent simulations with the same parameters) are represented respectively as dots and shaded areas, while the exact values are shown with dashed lines.
Panel $b$ shows that accurate results are obtained for the three parameters over a wide range of growth rates.
Panel $c$ shows that  accuracy improves as the filtering bandwidth increases.
Panel $d$ shows that accurate results are obtained for signals as short as approximately $T \simeq 50-100$~s (to be compared to the acoustic period $2\pi/\omega_0 \simeq 7$~ms and the characteristic time $1/|\nu| \simeq 170$~ms in this case).
The larger spread of $\kappa$ observed for large negative growth rates  results from the loss of statistical accuracy: in this range of $\nu$ the system remains mostly in a regime of low-amplitude oscillations and the nonlinearity is seldom activated.

\bibliographystyle{plain}
\bibliography{AFPE}

\end{document}